# BABEL OF UNITS

## THE EVOLUTION OF UNITS SYSTEMS IN CLASSICAL ELECTROMAGNETISM

Neal J. Carron

Rock West Solutions, Inc.
Santa Barbara, CA 93101

njc04@cox.net

May 21, 2015

# BABEL OF UNITS

## CONTENTS

# 1   INTRODUCTION

Few things are more confusing than the various sets of units employed in classical electromagnetism[1]. Five systems (of very many[2]) in common use are:
- The International System of Units (SI)
- Gaussian
- Variant-Gaussian
- Heaviside-Lorentz (rationalized Gaussian), and
- Various sets of "Natural" units.

The International System of Units, universally abbreviated SI (from the French *Le Système International d'Unités*), is the modern metric system of measurement. It applies to all of physics, and is now officially preferred. In electromagnetic context it used to be called MKS or MKSA, or, more precisely, Rationalized MKSA.

Gaussian is commonly used in electromagnetism, especially in fundamental discussions. It has replaced pure esu or emu. SI and Gaussian are by far the most common systems today.

Variant-Gaussian[3] is sometimes preferred to Gaussian because of the size of its current unit. It has no standard name, and is sometimes confusingly also called Gaussian.

Heaviside-Lorentz and natural units ($\hbar = c = 1$) are commonly used in Quantum Electrodynamics, less so in classical electrodynamics[4].

Other natural systems are often employed (e.g. *Atomic Units* in atomic structure studies). Natural units systems are not really for classical electromagnetism, and are not discussed here. A good discussion is in Klauber [Kl13]. See also Tomilin [To99].

For most purposes one need understand only the two major systems, Gaussian and SI. The surest way to understand the Gaussian is to understand the two earlier systems of which it is an amalgam: pure Electrostatic Units (esu), and pure Electromagnetic Units (emu). Each of these latter two systems is essentially never used anymore. However the letters "esu" are sometimes still loosely used when referring to electric quantities in Gaussian units, such as electric field (1 esu = $3\times10^4$ V/m) or charge (electron charge = $4.803\times10^{-10}$ esu).

---

[1]  One is the units in radiometry and photometry. Nit, stilb, Talbot, radiant exitance, Lumen, candela, lux, Lambert, phot, sterance, apostilb, and the Bougie-Hectomètre-Carrè are among the quantities that occur. See the IR Handbook, Section 1.1.

[2]  A few examples: Sommerfeld [So52] discusses a five dimension system in which the magnetic pole is the fifth basic quantity, but does not use it. The Russian Kalantaroff proposed a system with four basic dimensions: length, time, charge, and magnetic flux [Ka29. So52]. In that system mass would be expressed in terms of his four basic dimensions; see [Ki62]. Ampere himself used his own system, called *Electrodynamic Units*. Cohn (1900, 1927) introduced another system [So52]. Mendelson [Me15] discusses one early suggestion. Many others, some bizarre, have been proposed.

[3]  This system appears to have no official name. We simply call it Variant-Gaussian.

[4]  One of the earliest uses of units with $c = 1$ is in Einstein's 1934 lecture, well before quantum field theory matured [To07].

Here we recount the development of the six systems: *Electromagnetic*, *Electrostatic*, *Gaussian*, *Variant-Gaussian*, *Heaviside-Lorentz*, and *SI*, and come to appreciate why there is more than one system in the first place.

Handbooks and texts provide conversion factors and some technical discussion, but not always a good understanding of how so many unit systems came to be (with good reason; it's a long story). Several texts provide good discussions (Jackson, Panofsky & Phillips, Stratton, Sommerfeld, etc.). Jackson's treatment is straightforward and, to the present writer, the clearest of textbook discussions. He suggests Birge [Bi34, Bi35a,b]; these three papers are indeed excellent sources of understanding, emphasizing the arbitrariness in the *number* of base dimensions and the *choice* of units. They were written during a time (1930's) of vehement discussions about which system should be adopted internationally. See also Page [Pa70a].

The article by Silsbee [Si62] is a thorough comparative and historical treatment of the various systems, although it belabors the different conceptual approaches of theorists and experimentalists.

It is surprising that a thing ostensibly as simple as units can be so troublesome. What should be simple rules have turned out to be very confusing. The subject of units in electromagnetism has confounded a huge number of scientists for more than 180 years. Untold papers have been written advocating this or that system. References cited in the three papers by Birge from the 1930's allude to scores of them. In 1940 one physicist wrote, referring to the history of units during the preceding approximately 100 years,

> "Systems were conceived and forgotten with equal speed. Arguments arose as to whether *H* really did have the same dimensions as *B*, arguments which could only have arisen because no one knew what system he really was using. … So confused did the situation become that experts on the experimental aspects of magnetism published such statements as 'the dimensions of $\mu$ [magnetic permeability] are unknown'. … The International Electrotechnical Commission [I.E.C.] voted not on what system of units should be used but on whether $\mu$ really did or did not have dimensions." [Va40, p.225][5]

Alluding to the many systems recently in use, the book by Varner is entitled *The Fourteen Systems of Units* [Va48]; eight of the fourteen are electromagnetic, six are mechanical. Silsbee [Si62] opens his list of 122 references with the sentence, "The literature dealing with systems of units and dimensions is extremely voluminous". In addition to the present essay, books have been written just to explicate units [Varner, Va48; Vigoureux, Vi71; Cohen, Co01].

There are several international organizations, apparently all necessary, one of whose purposes is to review proposed units systems and approve them for official, national and international use. The important ones are

• *International Bureau of Weights and Measures* (BIPM, actually a central French laboratory near Paris, established 1875 following the international Treaty of the Meter in which 17 countries,

---

[5] The statement that the dimensions of $\mu_o$ and $\epsilon_o$ are unknown also appears in the 1922 physics encyclopedic dictionary by Glazebrook, Vol. 2 [Gl22], under "Electrical Measurements, Systems of", Section (1), p. 211. And see Birge's personal struggle with the matter [Bi34, p. 41].

including the U.S., approved the new physical standards for the kilogram and the meter), which is under the supervision of

- *International Committee for Weights and Measures* (CIPM), which itself is under the authority of
- *General Conference on Weights and Measures* (CGPM).
- *International Electrical Congress* (the first was 1881 in Paris), which in 1904 recommended the creation of the
- *International Electrotechnical Commission* (IEC) (founded 1906), which by 1923 had 10 technical committees, and by 1980 had 80.
- *British Association for the Advancement of Science* (founded 1831) is an older organization with a broader charter than establishing units, but instrumental in defining them, all with innumerable subcommittees on selected topics.

To understand electrodynamics it is not necessary to follow all the history, or even to understand units at the level we discuss here, so long as the system is properly defined. In principle Jackson's treatment is all you need. However it is common to learn the subject first (and sometimes only) in SI units, leaving one ill equipped to read much literature written in Gaussian units. And when a student sees a definition "1 Ampere is that current in two parallel wires 1 m apart whose repulsive force is $2 \times 10^{-7}$ N/m", one can't help but wonder "where did $10^{-7}$ come from, and why 2?" When one sees that 1 Oersted (Oe), a unit for magnetic field *H* in Gaussian units, is related to SI units by 1 Oe = $1000/4\pi$ A/m one wonders how that comes about. There are reasons man-made things evolved the way they did (at least historical reasons, if not logical ones); we hope our discussion helps clarify the reasons. We fill in between the lines of textbook treatments, providing historical background material that may appeal more to curiosity than to necessity. In any case one ends up with a much more satisfying overall understanding.

We summarize relevant features of important systems in a colloquial, rather than pedantic, manner, partly by sketching their history, which we find helps understanding. This essay is intended to be tutorial; it should appeal to students desiring a fuller discussion than in most textbooks.

Conversion tables, not given here, can be found in many texts, along with summary tables of equations in various systems, e.g. Stratton [St41], Panofsky & Phillips [Pa62], Becker [Be64], Lorrain and Corson [Lo70], Jackson [Ja75], Smythe [Sm89], Griffiths [Gr13], and in the papers by Silsbee [Si62] or Desloge [De94].

First recall the difference between *units* and *dimensions*.

## 2  UNITS vs DIMENSIONS

*Dimensions* are the fundamental independent quantities of the theory. The primary, or base, dimensions are usually mass, length, and time (in SI, current is a fourth base dimension). Equally confusing as the plethora of units systems is the fact that the *number* of base dimensions is arbitrary. SI uses the four just mentioned. Emu, esu and the Gaussian systems use only three: mass, length, and time; current and charge are expressed in terms of them. Other systems, especially the so-called Natural Units, use fewer.

Sommerfeld stresses the arbitrariness in the number of basic dimensions [So52, §8]. Sometimes base dimensions are loosely (and improperly) referred to as base units.

***Units*** are convenient amounts of the fundamental quantities defined for use as reference, for example: gram for mass, centimeter for length, second for time (and, in SI, Ampere for current).

Other derived quantities (energy, magnetic field, conductivity, ... ) have dimensions belonging to the primary set of three or four, but, when convenient, have separately defined and named units, such as erg (g·cm$^2$/sec$^2$), Joule, Tesla, Siemens, Ohm, etc.

Units as such are a bit idealized. ***Standards*** are physical realizations of units. They evolve as technology evolves. The *standard* for the second *unit* of the *dimension* of time was, in the early 1800's, the period of a certain pendulum at 45° latitude. In 1832 Gauss replaced it by the mean solar second; that has since been supplanted by the number of oscillations of a certain optical line in Cs$^{133}$. The standard for the meter unit of the dimension of length was originally a carefully maintained Pt-Ir alloy bar in Paris. That standard was replaced by a certain number of wavelengths of a particular orange Kr$^{86}$ line; and that has now been replaced by the distance traveled by light in vacuum in a certain fraction of a second. The Ohm standard started as the resistance of a column of Hg, but has evolved to being defined by the quantum Hall effect.

## 3   THE PROBLEM

In SI current is a fundamental quantity, and its unit is the Ampere, defined in terms of force between two current-carrying wires. But in esu and the Gaussian system, the dimension of current is dyn$^{1/2}$·cm/s, and the unit of current is given the name "statAmpere" (1 stA = 1 dyn$^{1/2}$·cm/s). Its equivalent in SI is 1 stA ≈ 3.336×10$^{-10}$ Amperes. In emu and Variant-Gaussian, current has dimensions dyn$^{1/2}$, the unit used is the "abAmpere" (1 abA = 1 dyn$^{1/2}$). Its equivalent in SI is 1 abA = 10 Amperes.

In SI charge is not a fundamental quantity; it is simply current×time. Its unit is 1 Coulomb = 1 Ampere·sec. In electrostatic and electromagnetic units also, or in Gaussian units, charge is not a primary quantity. Rather, in esu and Gaussian the unit of charge is 1 statCoulomb = 1 dyn$^{1/2}$·cm, often called simply "1 esu". In emu the unit of charge is 1 abCoulomb = 1 abA·s = 1 dyn$^{1/2}$·s. In emu resistance has the dimensions of velocity; in esu or Gaussian it has the dimensions of 1/velocity; neither smacks much of resistance.

In SI the magnetic field $H$ has dimensions A/m, and does not have a separately named unit; $H$ fields are stated in A/m. However, the magnetic induction, $B = \mu_o H$, which has dimension kg/(A·s$^2$) does have a separately named unit, the tesla. In emu or Gaussian units $H$ has dimension of stC/cm$^2$ = (g/cm)$^{1/2}$s$^{-1}$ = abA/cm; the unit in which $H$ is often specified is the oersted; 1 Oe = 1 abA/cm, SI equivalent is 1000/4π A/m (= 79.58 A/m). However, in vacuum, $B = H$ in these systems, yet $B$ is given a separately named unit, the gauss.

This makes it clear that it is confusing. Sommerfeld says, "We have frightened generations of students with these two sets of values for charge and field strength ... ", referring to esu and emu [So52, p. 54]. He then proceeds to use (the then new) SI in the book[6] (actually Giorgi's MKSQ system, essentially the same as SI). Conscientious students have such difficulty with it that one lecturer attempts to present Maxwell's Equations in a form independent of units [Vr04]. This might confuse students more.

While international committees are no doubt diligent, one electrical engineer addressing the 1983 International Magnetics Conference in Philadelphia, and seemingly apologizing for the confusion of units, observed

> " … International committees arrive at their decisions by the same irrational procedures as do various IEEE committees that you have served on." [Br84]

The distinction between ***B*** and ***H*** was historically a particularly confounding point, and to many it still is. In the pioneer days, Maxwell was ambivalent about whether there was a conceptual difference between the two magnetic fields [Ro00]. The same reference reports

> "Richard Becker (1887–1955) wrote in 1932 that 'the distinction in principle between *D* and *E* [and between *B* and *H*] ...in empty space...has been absolutely abandoned in modern physics.' The distinction had not been abandoned, however. In that same year, in the course of an informal meeting of British and Continental physicists in Paris in July, Sir Richard Glazebrook [1854-1935]
>
>> referred to the fact that he was one of the last surviving pupils of Maxwell and he felt convinced from recollections of Maxwell's teaching that [Maxwell] was of the opinion that *B* and *H* were quantities of a different kind. When a vote was taken nine were in favor of treating *B* and *H* as quantities of a different nature, whilst three were in favor of regarding *B* and *H* as quantities of the same nature.
>
> Four years later, in 1936, a subcommittee of the International Electrotechnical Commission (IEC) proposed the names 'gauss' and 'oersted,' respectively, for the cgs electromagnetic units of *B* and *H*, respectively, even though *B* and *H* in a vacuum have the same numerical values and dimensions in that system."[7]

Furthermore, one finds apparently contradictory statements in the literature. The highly respected text by Stratton [St41, p. 17] states "… all electromagnetic quantities cannot be expressed in terms of these three [length, mass, time] alone". While 50 years earlier the founding father Maxwell stated [Ma91, Art 620] "Every electromagnetic quantity may be defined with reference to the fundamental units of Length, Mass, and Time." Can unit systems be so complicated as to give rise to such misunderstanding? Birge [Bi34] alludes to other inconsistencies and confusion in literature of the early 20th century.

---

[6] Sommerfeld was one of the most prominent physicists in the first half of the 20th century; his students included Werner Heisenberg, Wolfgang Pauli, Hans Bethe, Peter Debye, Walter Heitler, and others. Yet a modern commentary by Chambers [Ch99], referring to a 1950 committee's report on teaching physics, states "it [the committee's report] acts as an antidote to the highly misleading book by Sommerfeld, which has been responsible for a great deal of confusion." When it comes to systems of units in electromagnetism, it appears there are no unquestioned authorities.

[7] Griffiths [Gr32] reminds us that Maxwell defines the relation between *B* and *H* in two ways. In the first way, *B* and *H* have the same dimensions, and in any proportionality such as $B = \mu_o H$, the constant $\mu_o$ would be dimensionless. In the second way, *B* and *H* have different dimensions, and the constant would have dimensions. Griffiths relates the same 9 to 3 vote.

Despite some 180 years to get it right, the subject still produces argumentative literature, as evidenced, for example, by perusing the pages of relevant journals such as the *American Journal of Physics* or *Metrologia* over the past recent years.

## 4 OVERVIEW

The different systems of electromagnetic units arose historically from different conventions of relating electromagnetic quantities (charge, current, fields, ..) to mechanical quantities (length, mass, force, energy, ...). The need for such relations arose in the 19$^{th}$ century with the rise of electricity and magnetism after centuries of mechanics. Mechanics involved only the fundamental quantities mass, length, and time, and the dynamical quantities force, energy, momentum, ... . But when quantitative investigations by Ampere, Coulomb, etc., revealed phenomena besides gravity and springs that could produce forces, one needed to know the force between two charges or between two current-carrying wires, and so, to quantify the relationship, measures and units were needed for charge and current.

### 4.1 AN INSTRUCTIVE IMAGINARY HISTORY; THE CASE OF GRAVITY

Although there was no such time, one can imagine a time at which Newton's laws of motion were known, but gravity had not yet been discovered. Horses, springs and muscles were the only known forces. Newtonian mechanics ($F = ma$) explained motion under those forces. The known fundamental quantities were length, time, and inertial mass. Then experimentalists noted that, in addition to exhibiting inertia, objects exerted a force on one another. Gravity was discovered. The force was observed to be proportional to the product of the masses $m_1$, $m_2$ and inversely to the square of their separation $r$:

$$f \propto \frac{m_1 m_2}{r^2} \tag{1}$$

To make this proportionality an equality a proportionality constant is introduced:

$$f = G \frac{m_1 m_2}{r^2} \tag{2}$$

The known dimensions of $f$, $m$, and $r$ mean the dimensions of $G$ must be force·length$^2$/mass$^2$. Its magnitude is determined from measurements. Now this new force fits into existing mechanics. The "source" of gravity is the mass $m$, the same as the inertial mass.

With hindsight today (based on experience with electrostatic forces) it would have made equal sense to consider the source to be a gravitational "charge" $g$, having nothing to do with inertial mass except that experiments showed $g$ proportional to $m$.

Then one would have written $f \propto g_1 g_2 / r^2$ for the force between two gravitating bodies. The corresponding equality would be $f = G' g_1 g_2 / r^2$ with $G'$ some new constant. As with $G$ and $m$, only the product $G' g_1 g_2$ enters the basic force law, and therefore *only the product $G' g_1 g_2$, or $G m_1 m_2$, can be measured*. At this stage, where we have discussed only forces, separating the product $G'g_1g_2$ into factors $G'$ and $g_1g_2$ is make-work, it contributing nothing useful.

However, from relations (1) and (2), one understands that the existence of the constant $G$ or $G'$ is a property of the underlying force, and that $g$ is a property of each gravitating body. In spite of the fact that measurements cannot separate them, one now has a conceptual reason to separate $G'$ from $g$. In addition, when an intermediary *field* is introduced (force per unit charge), of strength $f/g_2 = G'g_1/r^2$, dimensions and units must be chosen for both $g$ and $G'$. No principle of physics dictates or even suggests the "proper" way to do this. How the separation of $G'$ from $g$ is carried out is up to the physicist. It had to be; physicists were breaking new ground by trying to make sense out of the new gravity in terms of old mechanics. There were no rules.

Since the force *law* fixes only the *dimensions* of $G'g_1g_2$, once dimensions are chosen for $g$ the dimensions of $G'$ are (force)·(length)$^2$/$g^2$. The dimensions of $G'$ change if one chooses to change the dimensions of $g$. Likewise the *measured* force $f_{measured}$ determines the *magnitude* of the product $G'g_1g_2$, and dictates that if one chooses a magnitude for $g$, the magnitude of $G'$ is $r^2 f_{measured}/g_1g_2$. The magnitude of $G$ changes if one selects a different magnitude scale for $g$.

In reality, since the magnitude of the force was proportional to an object's inertial mass, the actual choice made was to use for $g$ the same dimension and unit for gravitational charge as already used for inertial mass, $g = m$. This gravitational charge was later known as gravitational mass to distinguish its role conceptually from inertial mass, the former being responsible for the force, the latter responsible for a particle's reaction. Since mass and its unit were already known, Eq (2) can be used to *define* the dimensions of $G$; the dimensions of $G$ are force·length$^2$/mass$^2$, purely in terms of existing dimensions. Its magnitude was empirically found to be $G = 6.674 \times 10^{-11}$ N·m$^2$/kg$^2$. Thus there was no need to introduce a new unit or "charge" in gravitational mechanics.

This imaginary historical procedure for the gravitational force is analogous to the actual historical procedure for the electrostatic force. Although primitive electrostatics was known since the ancients, that force became understood and quantified only well after Newton's laws of force were enunciated. One had to assign a parameter, called the electric charge $q$, to quantify the strength of the source of the force between two separated charges.

### 4.2 THE REAL (BUT VERY SHORT) HISTORY OF ELECTRICITY AND MAGNETISM

We present enough history to provide the background context during the period of time that unit systems were being developed. A fuller but still terse history is given, e.g., by Errede [Er11].

Electricity and magnetism have an ancient history. Lodestone was a natural mineral that attracted iron. The magnet got its name from the region known as Magnesia on the Aegean coast of NW Turkey, in which lodestone was found.

The Greek word ελεκτορ (elektor) means "beaming sun". Amber, fossilized tree sap, is a hard, pretty, goldish brown material that sparkles orange and yellow in sunlight. Hence the Greek word ελεκτρον (sometimes ηλεκτρο) for amber. Amber was found near the Baltic coast of Germany. When rubbed with wool or cloth it was observed to attract small particles. That property would wear off after a short time; otherwise it might have received as much attention as the permanent lodestone, and, by analogy, electricity might be known today as Balticity. But electricity won out over Balticity (and possibly amberism). Similarly, amber in Latin is *electrum*, and *electricus* means to produce from amber by friction. These words are the roots of present day terminology [Wh51].

During the 1800s it became evident that electric charge had a natural unit, which could not be subdivided any further, and in 1891 George Johnstone Stoney (1826-1911) proposed to name it "electron". When J.J. Thomson discovered the light particle which carried that charge, the name *electron* was applied to it.

Minor advances in magnetism occurred during the Middle Ages. In 1600, in attempting to explain the alignment of free floating lodestones toward the earth's poles, Gilbert (1544-1603) proposed that the earth itself was a huge magnet. Descartes (1596–1650) then tried to explain magnetic forces via his vortex model of the aether.[8]

Eventually quantitative measurements began. Measurements before the 1830's, although "quantitative", were only relative. There was, for example, no absolute measure of current. One could only determine that the force produced by current *a* was, say, 1.6 times the force produced by current *b*. The Ampere, or any other unit of current, did not exist.

Early scientific work was either in magnetostatics or in electrostatics; they were separate subjects. Even today, as long as electric and magnetic fields are not both involved in the same problem (i.e., in just electrostatics or in just magnetostatics) it is easy to understand units. Difficulty understanding units arises only when discussing electric and magnetic phenomena together (i.e., electromagnetism) in a single unit system.

The first hint that there was any connection between electricity and magnetism was found in 1820 by Oersted in Denmark[9], who discovered that an electric current in a wire (a purely electrical phenomenon) would deflect a nearby compass needle (a purely magnetic phenomenon). Biot (1774-1862), Savart (1791-1841), and Ampere (1775-1836) in France and Faraday (1791-1867) in England picked up on the discovery and extended it. In the 1820's Carl Friedrich Gauss (1777-1855) and Wilhelm Eduard Weber (1804-1891) in Germany were also investigating Oersted's discovery. But Gauss was primarily continuing his measurements of the geomagnetic field, important for maritime navigation. In 1832 his work on geomagnetism became a turning point in the history of units when he found a way to finally measure magnetic fields in an absolute sense. He showed how to define and measure a magnetic field in terms of the gram, centimeter, and second (his method is reproduced in Appendix A). Weber, working with Gauss, was soon able to extend Gauss' magnetic method to absolute measures of electric quantities as well.

---

[8] Maxwell gave the name "rotation" (abbr: rot) to the curl operation ([Ma91], art 25), very likely from early physical pictures of Descartes' rotating vortices, and hydrodynamic vortices, and from Maxwell's own hydrodynamic-like picture of the aether. And yet the term "curl" is also attributed to Maxwell [Ma71]. But it was J. W. Gibbs in the U.S. who introduced the familiar "dot" and "cross" notation for scalar and vector products of two vectors [Hu12]. Previously the scalar product was noted (*A*,*B*) or *S*(*A*,*B*) or sometimes just *AB*, and the vector product [*A*,*B*] or *V*(*A*,*B*). One still finds the term "rot" in older literature.

[9] Earlier hints were occasionally documented, e.g. Dod [Do35], reproduced in Appendix C. See also the next article in that journal [Co35].

In the physics community, effects of one body on another, be they gravitational, electric, or magnetic, were still thought of as instantaneous "action-at-a-distance" forces. Most physicists found action-at-a-distance repugnant, but could think of no viable alternative. Finally, in the 1840's, Faraday introduced "lines of force" in magnetism, and eventually the concept of a *field* in physics. The field was to be a dynamical, physically real quantity itself, and would eliminate the distasteful action-at-a-distance concept, although the idea was not rapidly accepted by other physicists. Even then, a medium was needed to transmit field effects; in vacuum that medium was called the ether, some mysterious stuff pervading all space.

Building on work by Faraday, Ampere, and others, in 1873 Maxwell (Scottish, 1831-1879) developed a mathematical theory of the field, and wrote a rambling and difficult[10] two volume *Treatise on Electricity and Magnetism*. He had introduced the displacement current, the seminal addition that led to his famous equations, earlier in 1861.

However his four famous vector equations do not appear in the *Treatise*. Heaviside, who studied it at length, said they were in there in a non-obvious "latent" sense, but not "patently". Rather there is a set of twelve principal equations, lettered A through L. The vector and scalar potentials, rather than the fields, figure prominently in them.

Maxwell died in November, 1879 at the age of 48, when Einstein was eight months old, after completing only one quarter of the $2^{nd}$ edition of his *Treatise*. By that time Maxwell "had convinced only a very few of his fellow countrymen and none of his continental colleagues" [Hu91, p. 1] that his theoretical unification was correct. It was left to his followers to complete that edition, published in 1881 and edited by W. D. Niven[11].

Maxwell's disciples had to extract from the first edition a coherent, understandable exposition of his theory. Those individuals were, principally, George F. FitzGerald (Irish, 1851-1901, of "contraction" fame), Oliver Lodge (British, 1851-1940), and Oliver Heaviside (British, 1850-1925). O'Hara & Pricha [OH87] and Hunt [Hu91] refer to them as *The Maxwellians*. Heinrich Hertz (German, 1857-1894) made later significant contributions. Apparently these followers were often led astray by obscurities, errors, and ambiguities in Maxwell's book. "Extracting a consistent and comprehensive theory from Maxwell's book was among the greatest and most difficult of the Maxwellians' tasks" [Hu91, p. 34]. As Hunt put it, it was these gentlemen who transformed Maxwell's equations into "Maxwell's Equations".

Much work went into devising mechanical models of the ether; Maxwell had a model, Lodge had his, and FitzGerald had his. Every physicist had one or several pet models. Conceptual models of the ether go back to John Bernoulli (1710-1790) and Leonhard Euler (1707 – 1783) [Wh51], but "serious" mechanical models were devised in the $19^{th}$ century. Scientists argued about it endlessly. It was also thought that, while light is electromagnetic in nature, light could not be *created* electromagnetically, and it was not realized that a time-varying current would generate waves. No one thought about creating waves of lower frequency than light until years later.

---

[10] The words are those of L. Pearce Williams who wrote the Foreword to Bruce Hunt's *The Maxwellians* [Hu91].
[11] Sir William Davidson Niven (1842–1917), Scottish born influential mathematician and physicist. He also edited Maxwell's Scientific Papers.

In the early 1880's, a passage in Lord Rayleigh's *Theory of Sound* led FitzGerald to realize that an oscillating current *would* radiate electromagnetic waves similar to light. This realization opened up the full frequency spectrum, and made it clear that, since oscillating currents occur often in ordinary electrical devices of the day, electromagnetic waves of various frequencies are all around us. Later FitzGerald used Rayleigh's retarded potentials[12] to clarify the propagation of a wave from its generating location to distant points at the speed of light. By 1883 FitzGerald realized that discharging a capacitor in a capacitor-inductor-resistor circuit could produce waves in practice, with wavelengths of a few meters.

Detecting those waves was another matter. FitzGerald thought of using a resonant circuit, but he determined that his design wouldn't work.

The theory of electromagnetic waves was barely sketched in Maxwell's *Treatise*, but by 1883 the Maxwellians had advanced it to a soundly based theory. By the mid 1880's Heaviside had simplified Maxwell's twelve equations to the now standard vector four, and by the mid 1890's they became accepted as the standard form.

England was connected to the continent in 1851 by the world's first undersea telegraph cable, from Dover to Calais. Telegraphy was the latest electrical technology (along with electrochemistry and battery technology), and England saw its potential for commerce and news and continuing world leadership. The first *successful* trans-Atlantic cable was laid in 1866; others followed world-wide in quick succession. Nearly 100,000 miles of cable were in place by 1885. World-wide messaging now took minutes instead of days and weeks.

Oliver Heaviside began his career as a telegrapher, working in 1868 on an Anglo-Danish cable. The design and improvement of cables was a merging point of basic science and practical technology. Over the years he applied his increasing understanding of Maxwell's theory to problems of telegraphy, especially to undersea cables, and significantly developed their theoretical underpinnings. For example, Lord Kelvin had calculated how a signal propagated down a cable, but ignored the displacement current and concluded that the voltage and current *diffused* along the cable. Heaviside emphasized the importance of the neglected term and developed the transmission line equations in much their modern form. All this work led to improved understanding of electromagnetic waves in general. However it was still believed that in vacuum the underlying "mechanism" was mechanical; hence continued emphasis on mechanical models of the ether.

Working with Maxwell's theory, freshly elaborated by the Maxwellians, in 1884 J. R. Poynting (British, 1852-1914) answered a long standing problem of how energy got from a radiating current to another point in space. He was able to derive from Maxwell's Equations the law by which energy is transferred in electromagnetic fields, the now famous Poynting's Theorem, and the Poynting Vector [Po84][13]. His work provided much insight to later physicists investigating how to demonstrate the existence of the predicted waves.

---

[12] Heaviside preferred to call them "progressive" potentials [Hu91, p. 43]

[13] Heaviside made the same observations independently, shortly after Poynting [Hu12].

"When the 1880's began, Maxwell's theory was virtually a trackless jungle. By the second half of the decade, guided by the principle of energy flow, Poynting, FitzGerald, and above all Heaviside had succeeded in taming and pruning that jungle and in rendering it almost civilized." [Hu91, p. 128]

In 1888, working in Germany, Heinrich Hertz was able to produce electromagnetic waves using an induction coil to achieve high voltages, a capacitor to create a resonant circuit, and a spark gap which would fire at the resonant frequency. The receiver was a single loop of Cu wire, with nearly the same resonant frequency, and a small (~ 0.01 mm) gap between the ends. The receiver spark would discharge when the transmitter spark did, demonstrating transmission of electromagnetic energy through vacuum (air). Had Hertz not died in January 1894 at age 36 there is little doubt he would have gone on to further momentous discoveries.

The third edition of Maxwell's *Treatise* was published in 1891, edited by J. J. Thomson. None of the names FitzGerald, Heaviside, Hertz, Lodge, or Poynting occurs in the index (in the 1954 Dover reprint), while the earlier researchers, Ampere, Cavendish, Coulomb, Faraday, Gauss, Ohm, William Thomson (Lord Kelvin) and others do[14].

British scientists seemed to love their mechanical models of the ether[15], their counterparts on the continent less so. However, in spite of their emphasis at the time in basic electrodynamics, mechanical models of the ether actually played little part in the development of units. Over the years, as the field equations proved themselves, less reliance was placed on models. In 1884-1885 Heaviside, having toyed with and then discarded Quaternion methods, introduced vector notation in the equations, greatly simplifying them into what we now know as "Maxwell's Equations" [Hu91, Hu12].

An important contribution by him was to reduce Maxwell's original emphasis on potentials to equations directly in terms of the fields, with which we are familiar today, especially rewriting Faraday's law in its present day vector form in *E* and *B*. In the first edition of his Treatise (1873) Maxwell stated that the vector potential was "the fundamental quantity in the theory of electromagnetism and invested it with great physical significance" [Hu91, p. 116]. Rather, Heaviside recognized *E* and *H* as the quantities of real physical interest (that it is better to claim *E* and *B* as basic was realized only later, see Section 5.4) . FitzGerald eventually turned away from the potentials altogether, and, agreeing with Heaviside, came to denounce their use "an analytical juggle" that obscured the real physics of how energy was localized and propagated in an electromagnetic field [Hu91, pp. 116, 118].

Ultimately, in 1905, Einstein did away with the ether altogether, replacing it with ⋯ nothing. The new view, that electromagnetic waves didn't need any medium to support them, could be as difficult to adapt to as a mechanical ether, and took a long time to be accepted.

---

[14] Hunt mentions that Thomson was a supporter of the earlier Maxwell, and still favored the potentials over fields. One wonders if that influenced his omitting the younger physicists. [Hu91, p. 202ff]

[15] "Heaviside affirmed many times that there must be a mechanical ether; to deny it, he said, would be 'thoroughly anti-Newtonian, anti-Faradaic, and anti-Maxwellian' " [Hu91, p. 104]

To obtain the fundamental equations in matter (for the macroscopic average field) the electric polarization ***P*** and magnetization ***M*** must be accounted for. This was done with the introduction of the auxiliary fields ***D*** and ***H***. Discussion of these fields is presented below in Section 5.4.

### 4.3  HISTORY OF DEVELOPMENT OF UNITS SYSTEMS

As distinct from basic electromagnetic theory, there is a separate, but of course closely intertwined, history of units in that theory. Here we briefly recount the story, gleaned from many sources, of how fumbling beginnings evolved into the stable systems of today.

The beginnings of standardized *mechanical* units arose in the 18th century from a wild mix of units for weights and measures that differed from country to country and even from town to town[16]. Influential scientists (Lavoisier[17], Borda, Laplace, …) forced standardized units into the new stable government in France following the French Revolution (1790's).

The meter as the standard unit of length and the kilogram as the standard unit of mass were established in 1799[18], and the centimeter and gram were soon accepted. In 1832 Gauss argued that the time unit, 1 mean solar second, obtained from astronomical observations, be added as the standard unit of time (replacing an older period of a pendulum). Thus was born the MKS system (and the cgs system) of mechanical units, the early beginnings of the full SI[19, 20]. With such a system accepted internationally, a scientist could report, say, that he measured a force of $X$ gram·cm/sec$^2$, and be assured any reader could reproduce his work and make a meaningful, quantitative comparison, since everyone knew what a gram, cm, and sec were.

---

[16] Thus the *Troy Ounce*, still used in the precious metals industry, is derived from the original (17th-18th century) ounce unit used in the town of Troyes, France, a major trade route stop-over (some authorities question this story: see Wikipedia, *Troy Ounce*). The Troy ounce was itself a derivative from old Roman weight units. Other measures still in use at this time were traceable to Charlemagne (8th - 9th century).

[17] In the political turmoil, Lavoisier, alas, was executed by guillotine in 1794 at the age of 50.

[18] The meter was defined as one ten-millionth the distance from the north pole to the equator along the meridian through Paris. The dramatic story of the French expedition to measure the mountains and valleys along the path is told in *The Measure of All Things* by Ken Alder, including deceitful cover ups by otherwise reputable scientists. The citation on an award won by the book reads "Alder's new book is extraordinary geodetic soap opera that deftly combines gripping narrative, a vivid sense of place and local culture, and a very human exploration of the meaning, moral significance, and profound personal costs associated with the Enlightenment's embrace of measurement, precision, and rational standards."

[19] Not all countries updated quickly. Naughtin [Na09] reports that in 1838 "A survey of Switzerland found that the Swiss foot had 37 different regional lengths and the Swiss ell had 68 different lengths. There were 83 different Swiss measures for dry grain, 70 different measures for liquids, and 63 different measures for weights."

[20] Naughtin [Na09] reports that in the 1850's "William Thomson [Lord Kelvin] played a major part in creating the international system of units used in the world today. He was known to have called the English system of weights and measures 'barbarous.' In a lecture demonstration with a muzzle-loader rifle confusion between the avoirdupois dram (about 1.8 grams) and the apothecary's dram (about 3.9 grams) caused a student to put into Thomson's muzzle-loader more than twice as much gunpowder as he should have. This could have blown Thomson's head off. However, Thomson's finicky attention to details led him to check the amount with the student before the demonstration." We never learn. In 1999 the U.S. lost an expensive Mars mission by inconsistently using English and metric force units [NA99].

But there were no dimensions or units for any electric or magnetic quantities. Coulomb published his historic force measurements in 1784. His observations were only on relative amounts and proportionalities, and even up to 1830 there was no way to state how much charge he worked with.

Early experiments in electrostatics and magnetostatics employed Leyden jars, Voltaic piles (the first voltage source), and crude torsional balances, electroscopes, and galvanometers, with measurements always being only of relative magnitudes[21]. One could only compare the amount of this charge relative to some other charge.

In the early 1830's Gauss was continuing his measurements of the geomagnetic field, and wanted an absolute measure of its magnitude. He recognized the value it would be to science if one existed. In 1832 he devised a way to do just that, that is, to express the field magnitude in terms of cm, gram, and seconds (presented in Appendix A). This was a monumental breakthrough. For the first time in history scientists could express magnetic quantities in terms of widely known mechanical measures. Gauss presented his method to the Göttingen Scientific Society in 1832, but it wasn't published in paper form until 1841 (a transcript in German of the Society presentation appeared in 1833 [Ga33]). Gauss based his original theoretical development on the force $f = p_1 p_2 / r^2$ between two (artificial) magnetic monopoles $p_1$ and $p_2$[22]. But in 1833 he pointed out that Ampere's force [Eq (4) below] could be equated with its mechanical equivalent too; modern discussions of Gauss' units system now derive it from Ampere's Law[23].

At the time, Weber was working with Gauss, investigating Oersted's new discovery. He found that Gauss' method for an absolute measure of the magnetic field could be extended to electrical measurements as well. Together, by the 1850's, they devised absolute units applicable to electromagnetic phenomena in general. Their units system had no particular name; it seems to have been referred to as just the "Gauss-Weber method".

In laboratory work the useful quantities in electrical circuits are potential and current, and their ratio resistance. A potential of order 1 V[24], and a current of order 1 A are convenient reference values. Ohm's Law was published in 1827. It was a very valuable relation between current and potential for laboratory work. Unfortunately, the Gauss-Weber system had units of potential (equivalent to $10^{-8}$ Volt) and resistance ($10^{-9}$ Ω) that were not convenient for laboratory work, especially since all calculations were carried out by hand. And in the Gauss-Weber system resistance had the unintuitive dimensions of a velocity.

---

[21] Cavendish (1731-1810) measured current by "the level of personal discomfort".

[22] Forces between magnetic monopoles are essentially never discussed today. But in Gauss' time it was common. The $1/r^2$ behavior between two magnetic poles applies when the opposite poles of two long thin dipole magnets are far enough away to have negligible effect. It appears it was easier to confirm the Coulombic $1/r^2$ behavior between magnetic poles than between two electric charges, because the magnetic poles at the ends of magnets maintain their strength while electric charges are affected by leakage currents in air and imperfect dielectrics [Co01, p. 68].

[23] A simple demonstration of the equivalence of the magnetic pole and Ampere's Law derivations of Gauss' system can be found in Assis [As03, Sec. 3], or, preferably, in Golding [Go49].

[24] The Daniell cell, a widely-used battery voltage source at the time, produced about 1.07 Volts. It was supplanted barely 10 years later by the more stable zinc-mercury standard Clark cell, about 1.434 Volts.

In 1861 the British Association for the Advancement of Science (BAAS, or just B.A.) convened a committee to investigate standards for resistance. Maxwell, William Thomson (later Lord Kelvin), and Joule were principal members. The committee put forward the "B.A. unit" for resistance, to be $10^9$ units of resistance in the Gauss-Weber system, alternately known as the "Ohmad", which name soon became the "Ohm" [Ba61]. In the process the members introduced for the first time the concept of a "system" of units; previously there was no coherent, widely used set of units encompassing more than just a few phenomena. The committee noted the existing Gauss-Weber units system, and formalized it. It was named **The Electromagnetic Units System (emu)**, officially defined in the committee's report of 1873.

The committee then noted that an alternate, equally valid, system could be based on the Coulomb force $f = q_1 q_2/r^2$ between two charges, instead of the Coulomb-like force $f = p_1 p_2/r^2$ between two magnetic monopoles (or instead of Ampere's force between two wires). The committee called this system **The Electrostatic Units System (esu)**, also officially defined in the 1873 report[25]. Thus, while Gauss and Weber really "invented" emu, no one comparably invented esu; rather, it was a side development by a committee. Its unit for potential (equivalent to 300 Volt), current ($3.3 \times 10^{-10}$ A), and resistance ($9 \times 10^{11}$ Ω) were also inconvenient for laboratory work.

The shortcomings of emu and esu in the laboratory led applied physicists and engineers to seek a new, more practical set of units. The electromagnetic unit system was of greater value than esu to practical physicists working with the telegraph, largely because of the size of its current unit (equivalent to 10 Ampere), so the new system started with emu. That system's small resistance and voltage units could be improved if a large unit of length and a small unit of mass were used instead of the centimeter and gram [Ma91, Art 629; Si62]. Hence in a new practical system the length unit was taken to be the *quadrant* of the earth circumference ($10^7$ m), and the unit of mass was $10^{-11}$ grams (the *eleventh*-gram[26]). The unit of time remained 1 sec. It was called the QES (quad-eleventhgram-second) system[27]. In it, the unit of potential was called 1 Volt, current 1 Ampere, and resistance 1 "B.A. Unit", or later the Ohm, each very nearly (but not exactly) the same as today. The Farad as a unit of capacitance and the Coulomb as a unit of charge were introduced in this system[28]. Given the size of its mechanical units, clearly only the electromagnetic quantities, not mechanical ones, were "practical".

---

[25] The name *electrostatic units* system comes from its starting point, the electrostatic Coulomb Law, not from any restriction on its validity. The system is for all electromagnetic phenomena, just as emu is.

[26] In 1870 George Johnstone Stoney (FitzGerald's uncle) proposed a naming system for numerical power-of-ten prefixes. For a negative power, like $10^{-11}$, the prefix is to be the exponent followed by "th", and precedes the unit, thus $10^{-11}$ grams is read as 1 *eleventh-gram*. For a positive power, the prefix is replaced by the exponent following the unit, thus $10^9$ grams is read as 1 *gram-nine*. The quadrant would be 1 *meter-seven*. His suggestion was used for a while, but did not catch on for long. Rather, the naming system … micro, milli, … kilo, mega, … won out over Stoney's suggestion [Si62, fn 13].

[27] In 1889 an international electrical congress at Paris adopted the word *quadrant* as the practical unit of inductance (equivalent to $10^9$ cm in emu) [Pe95]. Four years later the international electrical congress of Chicago changed the name of the inductance unit to *Henry*, in honor of the American physicist Joseph Henry [ET00]. The name of the length unit apparently remained the *quad*.

> Two electrical engineers, Sir Charles Bright and Latimer Clark, oversaw the laying of the first transatlantic telegraph cable in 1858. They too advocated rigorous units standards, presenting proposals to the BA committee, and instituted the practice of naming units of physical quantities after prominent scientists.

In this practical system, the Ampere was defined by a physical standard[29]: 1 A was defined as that current which, flowing through a water solution of $AgNO_3$, deposited silver at a rate of 1.118 mg per sec. 1 Ohm was defined as the resistance of a column of Hg 106.300 cm long and 1 $mm^2$ in cross section, at a specified temperature. This was the "B.A. Unit" mentioned earlier, and was supposed to be a realization of $10^9$ units of resistance of the emu system. 1 Volt was the product of the two. These units for practical quantities, based on physical standards, were declared official by the International Electrical Congress in Paris in 1889, and remained official until 1948.

But slight discrepancies emerged. The practical Ohm just defined turned out to be 0.05% larger than the emu Ohm, and the practical Ampere was about 0.015% smaller than 0.1 abAmpere [Ko86]. The name "International Units" (not to be confused with *Système International*, *SI*) became associated with the units based on physical standards, and the name "Absolute Units" with those defined in terms of mass, length, and time as in emu. Today's units are the absolute ones. Standards are developed to match the unit; the unit is not defined by a standard. And today's standards evolve as technology evolves to provide the most accurate and reproducible methods.

Thus, the Electrostatic Units system itself, and the name Electromagnetic Units for the Gauss-Weber system, evolved from the 1862 committee work. And the QES system was active at the time. Maxwell's Treatise [3rd ed., Ma91] makes use of all three. From the 1860's to the early twentieth century emu was the most frequently used unit system.

One might think things had settled down, and that most physicists would have adopted at least one of the most recent systems. However, in 1881 "there were in various countries[30],

---

[28] Earlier, in 1861, Sir Charles Bright and J. Latimer Clark suggested the name Farad as a unit of charge, the Volt as a unit of resistance, and the Galvat as a unit of current [Ba61]. And at various times the name Weber appears to have been used to mean charge, current, magnetic pole, and magnetic flux density, before it settled down to magnetic flux. And "quadrant", "quad", and "sec-ohm" were suggested as the name for unit of inductance before "Henry" was adopted. These are other examples of the wild suggestions of units, names, and dimensions that occurred before stable systems and nomenclature were established. The names sturgeon, the Kelvin (for kilowatt), the thom (for magnetic induction), the gilb, the mac, and the mho (suggested by Kelvin as the unit of conductivity) were suggested and abandoned [Ba61, p. 538]. The mho was later reinstated and abandoned again in favor of Siemens.

[29] Why not keep the definition in terms of the Ampere force between two parallel wire currents? Because that force per unit length was extremely small and difficult to measure accurately. The force/cm repelling two wires 1 cm apart each carrying 1 Ampere is 0.02 dyn. 1 dyn is about the weight of 1 $mm^3$ of water. The physical standards were easier for worldwide laboratories to confirm [Ko86]. This is contrary to Gauss' original motivation (see Section 6.1).

[30] In 1933, referring to the historically slow pace with which countries adopted the metric system, Kennelly [Ke33] stated "… examination of the present … will show that this country [the U.S.] is in steady process of transition to the eventual complete adoption of the metric system, although it is not possible to predict the final official date of acceptance. … [estimating from historical experience] we might hazard the date of

12 different units of electromotive force, 10 different units of electric current, and 15 different units of resistance" [IEC 64].

In 1888 Hertz was testing Maxwell's prediction of electromagnetic waves. He noticed that a set of units more convenient than emu or esu for theoretical discussions, if not laboratory work, emerged if one specified electrical quantities in esu, and magnetic quantities in emu. Then fewer factors of *c* occurred in Maxwell's Equations (Eq (62) below, compared with (47) and (57)). Together with Helmholtz, he named this combined set the **Gaussian Units System**[31].

That system too had inconvenient units for potential (300 V), current ($3.3 \times 10^{-10}$ A), and resistance ($9 \times 10^{11}$ Ω), the same as esu. A still better (i.e., more convenient) system was sought.

Meanwhile, starting around 1892, Oliver Heaviside was ranting about the silly appearance of $4\pi$'s in the fundamental equations of electromagnetism that appear in emu, esu, and Gaussian units (Eqs (47), (57), and (62) below). The $4\pi$'s appeared as the result of choosing units so that the electrostatic force equation read

$$f = q_1 q_2 / r^2 .$$

with *f* in dyn and *r* in cm. That electrostatic equation was regarded as the fundamental starting point, and so units of *q* were chosen to make it simple (leading coefficient on the right-hand-side = 1). The electric field then reads $E = q/r^2$. But with that convention, Gauss' Law reads $\nabla \cdot E = 4\pi\rho$, containing an odd factor of $4\pi$.

It was much more rational, Heaviside said, to choose units for *q* so that the equation would read

$$f = q_1 q_2 / 4\pi r^2 ,$$

and the electric field would be $E = q/4\pi r^2$. That choice displays the fact that the total electric flux through a sphere, $4\pi r^2 E = q$, is the total charge inside. That would be a more rational choice than the result $4\pi r^2 E = 4\pi q$ coming from esu. A parallel argument applied to the magnetic flux from a magnetic pole. $4\pi$ should appear in the denominator of the force law for problems of spherical symmetry, and $2\pi$ for problems of cylindrical symmetry; they have no reason to appear in the fundamental field equations. Heaviside so emphasized the irrationality of the usual convention that his suggested approach actually came to be called *Rationalization* (or perhaps from the word *ratio*). Physicists nodded their heads, but were not energetic enough to embrace the suggestion and create a new system.

In 1901 the Italian physicist and electrical engineer Giovanni Giorgi devised a practical system reverting to the use of the meter, kilogram, and second, while retaining the Volt,

---

metric reform with us as, say, 1950; whereas, according to experience in calendar reform, it might be 1970." In 2015, 82 years after Kennelly's article, and 45 years after his most distant guess, it appears the U.S. still could care less.

[31] The usual practice today is to consider current an electrical quantity (a moving charge), and so measured in esu (statAmperes). If one considers current a magnetic quantity (source of the magnetic field), one would measure it in abAmperes, and one ends up with a variant of the Gaussian system; see Section 6.4.

Ampere, and Ohm units from the QES system. To do so he suggested that in addition to the usual meter, kilogram, and second, the unit of resistance, the Ohm, be a fourth primary unit in his system, the first attempt at a system with four primary dimensions, the fourth being an electromagnetic, non-mechanical quantity. Eventually the Coulomb replaced the Ohm as the fourth dimension (or fourth unit), and his system was temporarily called the Giorgi-MKSQ system (Q meaning charge, not quad). The MKSQ system eventually evolved into the modern day SI, in which the Ampere replaced the Coulomb as the fourth primary unit[32, 33]. By introducing an electromagnetic quantity as a fourth dimension, one could eliminate silly dimensions in earlier systems, in which the dimensions of current were $M^{1/2} L^{1/2} T^{-1}$, and resistance were $L/T$. Rationalization was built into the later versions of MKSQ and in SI. Jayson [Ja14] nicely tells the story of the development of the SI system. Newell [Ne14] summarizes its current status and plans to improve it.

Innovation is not always readily accepted. Kowalski [Ko86] states "… the physics community was deeply divided in the period of transitions to the SI and many physicists opposed the innovations. Their opposition was not always unjustified".

Electromagnetic Units were used for almost all work in electrodynamics between about 1850 and the 1890's, and either emu or Gaussian Units from 1890 to the 1940's, although Giorgi's MKSQ system was also employed (see Stratton [St41], Section 1.8). One book, by Gilbert [Gi32], was written entirely in emu. Stratton's 1941 book uses the rationalized Giorgi MKSQ system, really the same as SI except the Coulomb is the fourth unit instead of the Ampere. After the 1940's Giorgi's system, and later SI, were commonly used along with Gaussian.

It appears that esu, a perfectly valid system, was seldom if ever used as a general system for electromagnetism. It stands as a construct most famous for forming the basis of half of Gaussian Units. Because it underlies electric quantities in Gaussian Units, the three letter phrase "esu" is still used as a generic expression for electric quantities in Gaussian Units.

## 5   GENERAL RELATIONS IN ALL SYSTEMS

We now turn to the actual construction of a unit system.

For formal definitions of each system, the starting point is a force between two physical objects, such as two gravitating masses, two electric charges, or two currents (or, as was originally done by Gauss for derivation purposes, two magnetic monopoles). The physical object is considered the *source* of the force. One then chooses a parameter $q$ (a "charge") to quantify the strength of that object, and then inserts a constant $k$ in the force law to convert a proportionality to an equality. If the physical object is a point (electric charge), the force law takes the form of *Coulomb's Law*:

$$f = f_{\text{electrostatic}} = k_e \frac{q_1 q_2}{r^2} \qquad (3)$$

where $k_e$ is some proportionality constant. (No *a priori* principle requires the force be proportional to $1/r^2$; it is measured to be so in gravity and electrostatics). Eq (3) is a fundamental relation in electrostatics, and a defining connection between mechanics ($f$) and electricity ($q$). In older literature, e.g. [Bi34], $k_e$ in (3) is written as $1/\epsilon$, and in vacuum as $1/\epsilon_o$. At this stage of the discussion $k_e$, $\epsilon$, or $\epsilon_o$ are generic unit-dependent constants.

---

[32] The Ampere was added to SI officially only in 1954; the mole, and units for temperature (Kelvin) and photo-metric light intensity (Candela) were added later. There are seven primary quantities in the modern SI, four of them occurring in electromagnetism. The name *International System of Units* (*SI*) wasn't adopted until 1960 [NIST].

[33] Shortly before Giorgi's system, other practical systems had been suggested, usually also making use of the Volt, the Ampere, and the Ohm. Not everyone was pleased at the suggestion that "practical" units should replace the long-standing absolute emu system. In 1900, in the U.S., Fessenden wrote "It seems that there is a very strong feeling amongst electricians and electrical engineers of this and other countries in favor of a return to absolute units and the abandonment of the practical system … At present the whole system is in the greatest confusion" [Ba61, p. 538]. At that time the term "electrician" had a much broader connotation than today.

If the physical object is a line (long, fine wire carrying current $I$, with magnetic force between two), the force $f'$ is per unit length, and it takes the form of *Ampere's Force Law*:

$$f' = f'_{\text{magnetostatic}} = 2k_m \frac{I_1 I_2}{r} \qquad (4)$$

with $k_m$ here some other constant. (The "charge" in this case is a current). Eq (4) is a fundamental relation in magnetostatics, and a defining connection between mechanics ($f'$) and magnetism ($I$). The 2 is for later convenience[34].

In the case of gravity, dimensions and units for mass were already in common use, and, as stated, Eq (2) could then be used to define $G$. But in early electromagnetism $q$ and $I$ did not already have dimensions or units, so Eqs (3) or (4) could not be used to define $k_e$ or $k_m$ in terms of them. $k_e$ and $k_m$ must be chosen by the physicist. They had to be; physicists were breaking new ground by trying to make sense out of the new electric and magnetic stuff in terms of old mechanical stuff. There were no rules.

## 5.1 A RELATION BETWEEN $k_e$ AND $k_m$

As stated, Eqs (3) and (4) show that only the product $k_e q_1 q_2$ or $k_m I_1 I_2$ can be measured in mechanical terms. $q$ (or $I$) is a new non-mechanical quantity; to be chosen and specified. The different systems of units correspond to different *choices* of dimensions for $q$ and $I$, and $k_e$ and $k_m$. The freedom permitted in choosing $k_e$ and $k_m$, and another constant $k_B$ introduced in Eq (10), is the source of all the confusion in unit systems.

The ratio of Eq (3) to Eq (4) necessarily has the dimension of a length. And since charge is current times time,

$$[q] = [I \cdot t] \quad \text{(fundamental relation, essentially a definition)}, \qquad (5)$$

the ratio can be expressed, in terms of dimensions time (T) and length (L), as

$$\frac{f}{f'} \sim L \sim \frac{k_e}{k_m} \frac{q_1 q_2 / L^2}{I_1 I_2 / L} \sim \frac{k_e}{k_m} \frac{T^2}{L} \text{ , or}$$

$$1 \sim \frac{k_e}{k_m} \frac{T^2}{L^2}$$

(6)

---

[34] The origin of the factor 2 is explained in footnote 66, in the context of SI but applicable to all systems. As will be seen, in SI the proportionality constant is written $k_m = \mu_o/4\pi$ with $\mu_o$ a different constant. The $4\pi$ is to make SI a rationalized system.

This means that, with both equations (3) and (4) *expressed in a single unit system*, $k_e$ and $k_m$ cannot both have arbitrary dimensions. Rather, $k_e/k_m$ must have the dimension velocity$^2$. Eqs (3), (4), and (5) are starting equations, and impose this fundamental constraint when one chooses $k_e$ and $k_m$.

As in gravity, the force magnitude is measured empirically, determining the value of $kq_1q_2$. Separating the fixed units of $kq_1q_2$ into separate units for $k$ and $q$ is a matter of personal preference. Historically, different workers chose different units for $k$ or $q$, leading to different systems of units for electricity. (But, historically, only one system for gravity was ever introduced).

Both $f$ and $f'$, Eqs (3) and (4), can, of course, be measured. Measurements in the early 19[th] century showed the velocity $\sqrt{k_e/k_m}$ was very close to the then known speed of light. That the speed of light should pop out of electrostatics and magnetostatics provided a strong suggestion that light was electromagnetic in nature. We shall subsequently use

$$\sqrt{\frac{k_e}{k_m}} = c \text{ .}$$

(7)

Thus, given $c$, only one parameter, $k_e$ or $k_m$, is sufficient to establish dimensions of charge and current in terms of mass, length, and time (and perhaps a fourth dimension if one wishes, such as in SI). By minor extension one will obtain dimensions of potential, electric field, conductivity, etc.

With $\boldsymbol{j}$ the current density and $\rho$ the charge density, conservation of charge is expressed by

$$\nabla \cdot \boldsymbol{j} + \frac{\partial \rho}{\partial t} = 0 \text{ .}$$

(8)

## 5.2 THE MAGNETIC FIELD

The electric field is always defined as the force per unit charge,

$$E = \frac{f}{q} = k_e \frac{q}{r^2} \text{ ,}$$

(9)

which definition fixes its dimensions once one chooses $k_e$. But the magnetic field, at this stage of the discussion, is only known to be *proportional* to force/current, leaving another proportionality constant to be chosen. Indeed, the electric field $\boldsymbol{E}$ always enters the Lorentz force as $q\boldsymbol{E}$, but the magnetic field $\boldsymbol{B}$ may enter as $q\boldsymbol{\upsilon}\times\boldsymbol{B}$ or $q(\boldsymbol{\upsilon}/c)\times\boldsymbol{B}$ depending on the unit system. $\boldsymbol{E}$ can also be defined via the work done on a unit charge, but $\boldsymbol{B}$ cannot be defined that way, since it does no work. Rather, $B$ is simply proportional to the force per unit length on unit current, $f'/I_1$, at distance $r$ from another line current $I$. From (4) this defines $B$ as

$$B = k_B \frac{f'}{I_1} = 2k_B k_m \frac{I}{r} \qquad (10)$$

where $k_B$ is a new proportionality constant, again chosen by a human being and so differing from one system to another.

The dimensions and magnitude of $k_B$ are to be chosen; they then fix those of $B$. Since the dimension $[I/r] = [q/rt] = [cq/r^2] = [cE/k_e]$, we immediately see the dimensions of $B$ are

$$[B] = \frac{[k_B k_m]}{[k_e]}[cE] = [k_B][E/c] \qquad (11)$$

in which the right-most form has imposed constraint (7).

The Lorentz force on a charge $q$ with velocity $\upsilon$ is $F_L = q(E + \gamma \upsilon \times B)$, with $\gamma$ a constant depending on units. By requiring $\gamma \upsilon \times B$ have the dimensions of an electric field, and using (11), find $\gamma = k_e/c^2 k_B k_m = 1/k_B$, so the Lorentz force in any units is

$$F_L = q(E + \frac{1}{k_B} v \times B) \ . \qquad (12)$$

Thus, if one decides to build a unit system based on Coulomb's Law, Eq(3), one must choose dimensions and a numerical value for $k_e$, thus fixing the dimensions of charge, and the force between two charges. Then, from (7), $k_m$ will necessarily have the value $k_m = k_e/c^2$; Eq (4) then determines the force between currents. Finally one must choose a value for $k_B$, which fixes the dimensions of $B$, and (10) determines the magnitude of $B$ from a current.

Thus far we have mentioned only electrostatic and magnetostatic relations. Now consider time varying fields.

$E$ and $B$ are related not only dimensionally, but also dynamically by Faraday's Law, that the induced emf is *proportional* to the time rate of change of the magnetic flux[35, 36]. In modern differential form this law reads

---

[35] $B$ is called the magnetic induction vector because it is the field that determines electromagnetic induction via (13). This distinguishes it from $H$, which is known as the magnetic field intensity or the magnetic force. To some, $B$ is more properly called the magnetic flux density. However Faraday's "lines of magnetic force" are actually lines of $B$, not $H$. These names are most appropriate for historical or pedantic purposes; today, for a vacuum, one simply refers to $B$ or $H$ as the magnetic field. When discussing magnetic materials, one must distinguish $B$ from $H$.

[36] As recently as 2000 it was still unclear why the symbol $B$ was historically used for the magnetic induction [Ba00]. It may be an alphabetic accident. In his *Treatise* Maxwell listed nomenclature for quantities he discussed, in seemingly no fundamental order: first in the list was Electromagnetic momentum (= Vector Potential) $A$, second was magnetic induction $B$. Fourth in the list was the displacement vector $D$, and eighth was Magnetic Field Strength, assigned the letter $H$ [Ma91, Art 618]. It isn't clear whether he listed them in the alphabetic order of symbols already in use, or in some order deemed of dynamical significance. Vector notation was not used before Maxwell's time, so there may not have been any single symbol for vector dynamical quantities in those earlier days. However it is worth noting that, before Maxwell, $H$ was often used to denote the horizontal component of the geomagnetic field intensity; it is possible that later the symbol carried over to denote the full vector field.

$$\nabla \times \boldsymbol{E} + k_F \frac{\partial \boldsymbol{B}}{\partial t} = 0 . \tag{13}$$

Again one must choose the proportionality constant $k_F$. However, as for $k_e$ and $k_m$, dynamical equations impose a relation between $k_F$ and $k_B$. The relation can be revealed by considering the general Maxwell's Equations. We expand on Jackson's Appendix[37].

## 5.3 MAXWELL'S EQUATIONS IN ARBITRARY UNITS

The absence of magnetic monopoles, $\nabla \cdot \boldsymbol{B} = 0$, and Faraday's Law, Eq (13), are two of Maxwell's Equations.

The third equation, Gauss' Law, $\nabla \cdot \boldsymbol{E} = 4\pi k_e \rho$, is a direct consequence of the Coulomb force (3). Note from Eq (9) that for a point charge, $\boldsymbol{E} = k_e q \hat{\boldsymbol{r}} / r^2$. Taking its divergence and using

$$\nabla \cdot (\hat{\boldsymbol{r}} / r^2) = 4\pi \delta^3 (\boldsymbol{r}) ,$$

and $q\delta^3(\boldsymbol{r}) = \rho$, results in Gauss' Law valid for a point charge. Linear superposition then implies it is true in general.

The fourth equation is Ampere's Law, which, after Maxwell added the displacement current term proportional to $\partial \boldsymbol{E}/\partial t$, is

$$\nabla \times \boldsymbol{B} = 4\pi k_B k_m \boldsymbol{j} + \beta \frac{\partial \boldsymbol{E}}{\partial t} \tag{14}$$

The factor $k_B k_m$ comes from (10). The proportionality factor $\beta$ is determined by taking the divergence of (14) and using (8) and Gauss' Law,

$$0 = -4\pi k_B k_m \frac{\partial \rho}{\partial t} + \beta 4\pi k_e \frac{\partial \rho}{\partial t} , \tag{15}$$

showing that $\beta = k_B k_m / k_e$.

Collecting these results, we have Maxwell's Equations in the absence of material media including all factors $k_e$, $k_m$, $k_B$, and $k_F$, before any constraints on them; in this form they hold for any system of units:

$$\begin{aligned}
\nabla \cdot \boldsymbol{E} &= 4\pi k_e \rho & \text{(Gauss' Law)} \\
\nabla \times \boldsymbol{B} - \frac{k_m k_B}{k_e} \frac{\partial \boldsymbol{E}}{\partial t} &= 4\pi k_m k_B \boldsymbol{j} & \text{(Ampere's Law)} \\
\nabla \times \boldsymbol{E} + k_F \frac{\partial \boldsymbol{B}}{\partial t} &= 0 & \text{(Faraday's Law)} \\
\nabla \cdot \boldsymbol{B} &= 0 & \text{(Absence of magnetic poles)}
\end{aligned} \tag{16}$$

---

[37] The relation between our four constants and Jackson's $k_{1,2,3}$ and $\alpha$ is: $k_e = k_1$, $k_m = k_2$, $k_F = k_3$, $k_B = \alpha$.

In addition to constraint (7) between $k_e$ and $k_m$ arising from static forces, another relation, that constrains $k_B$ and $k_F$, can be derived from the two equations with time derivatives. In free space ($j = \rho = 0$), the two curl equations imply the wave equation:

$$\nabla^2 \mathbf{B} - k_F \frac{k_m k_B}{k_e} \frac{\partial^2 \mathbf{B}}{\partial t^2} = 0 \ . \tag{17}$$

Knowing waves propagate with speed $c$, the coefficients must obey

$$k_F k_B \frac{k_m}{k_e} = \frac{1}{c^2} \ , \tag{18}$$

which, together with (7) implies

$$k_F k_B = 1 \ . \tag{19}$$

Thus, with (7) and (19), only two of $k_e$, $k_m$, $k_B$, and $k_F$ are independent. Imposing those two conditions, and choosing $k_e$ and $k_B$ as the two surviving parameters, the equations become:

$$\begin{aligned}
\nabla \cdot \mathbf{E} &= 4\pi k_e \rho \\
\nabla \times \mathbf{B} - \frac{k_B}{c^2} \frac{\partial \mathbf{E}}{\partial t} &= 4\pi \frac{k_e k_B}{c^2} \mathbf{j} \\
\nabla \times \mathbf{E} + \frac{1}{k_B} \frac{\partial \mathbf{B}}{\partial t} &= 0 \\
\nabla \cdot \mathbf{B} &= 0
\end{aligned} \tag{20}$$

All unit systems[38] differ only in their choice of dimension and magnitude for $k_e$ and $k_B$.

We can now list the four parameters for the various systems (Jackson's Appendix Table 1; we've added Variant-Gaussian).

**Table 1. Electromagnetic Constants for Various Systems [Ja75]**

| System | $k_e$ | $k_m = k_e/c^2$ | $k_B$ | $k_F = 1/k_B$ |
|---|---|---|---|---|
| Electrostatic (esu) | 1 | $c^{-2}$ | 1 | 1 |
| Electromagnetic (emu) | $c^2$ | 1 | 1 | 1 |
| Gaussian | 1 | $c^{-2}$ | $c$ | $c^{-1}$ |
| *Variant Gaussian* | *1* | *1* | *See Section 6.4* | |
| Heaviside-Lorentz | $1/4\pi$ | $1/4\pi c^2$ | $c$ | $c^{-1}$ |
| SI | $1/4\pi\epsilon_o$ | $1/4\pi\epsilon_o c^2 \equiv \mu_o/4\pi$ | 1 | 1 |

---

[38] Except Variant-Gaussian. It is a hybrid, rogue convention, and, of all the systems discussed, is the only one that does not conform to the four parameter framework. See Section 6.4.

All systems are mechanically cgs except SI, which is MKS. SI is the only system with more than the three fundamental dimensions. The first four systems are unrationalized; only Heaviside-Lorentz and SI are rationalized (see Section 6.5). See [Wi15].

## 5.4 CONSTITUTIVE RELATIONS

When one wishes to extend the vacuum Maxwell's Equations to include behavior of macroscopic fields inside matter, one needs to account for the displacement of bound charges, and the existence of permanent and induced material magnetic fields.

Additional fields arise because of the *polarization* $P$ (electric dipole moment per unit volume) and the *magnetization* $M$ (magnetic dipole moment per unit volume) in a material[39]. $P$ and $M$ may exist in their own right, or be induced by external fields, or both.

It appears Kelvin was the first to suggest that two magnetic fields, now called $B$ and $H$, are useful in treating material media [Wh51, p. 219; Ke51, p. 269; Ke84, Section 517][40].

And evidently it was Oliver Heaviside who introduced the term *electric permittivity*, "indicating the capacity for permitting electric displacement", while Kelvin was the first to use the term *magnetic permeability*, in 1872, in analogy with fluid flow, as a measure of how readily magnetic fields permeate a medium, including vacuum [Ra00].

Jackson [Ja75, p. 189] emphasizes that $E$ and $B$ are the fundamental fields. In matter one accounts for $P$ and $M$ by defining the auxiliary fields $D$ and $H$ via the *constitutive relations*

$$D = \epsilon_o E + \lambda P$$
$$H = \frac{1}{\mu_o} B - \lambda' M \quad (21)$$

where $\epsilon_o$, $\mu_o$, $\lambda$, and $\lambda'$ are new constants, with different values in different unit systems (see Table 2 below)[41]. $D$ is the electric *displacement vector*[42]. $H$ is the *magnetic field intensity*.

The polarization $P$ and magnetization $M$ provide charge and current densities. The charge and current densities provided by electric polarization are

$$\rho_{pol} = -\nabla \cdot \boldsymbol{P}, \qquad \boldsymbol{j}_{pol} = \frac{\partial \boldsymbol{P}}{\partial t} \tag{22}$$

and the current density contributed by magnetization is

$$\boldsymbol{j}_{mag} = \nabla \times \boldsymbol{M} \ . \tag{23}$$

---

[39] As the term *polarization* could apply to electric or magnetic dipoles, the term *electrization* was once suggested by P. Cornelius to replace it. The idea did not catch on. When ambiguity is possible, one says *electric polarization* [Si62, p. 163b].

[40] The article by Roche [Ro00] is an invaluable aid when trying to understand Kelvin's original papers. It is also an excellent commentary on the "century-old controversy" surrounding the meanings of $B$ and $H$. Also worthwhile is the insightful and acerbic commentary by Chambers [Ch99].

[41] The forms in Eq (21) are conventional and (almost) universally agreed upon. The independent fellow Feynman chose a different way [Fe63]. Thanks to Dr. Steve Bottone for reminding the author of Feynman's unconventional convention. Actually, Coleman [Co73] also entertains Feynman's suggestion.

[42] So named in connection with the displacement of bound charges in dielectrics, though when introduced in 1861 it was not recognized that the charges were electrons bound to atoms. For the vacuum, Maxwell was one of many who had a dynamical picture of the aether, then considered a material medium. In it lines of $B$ or $H$ formed circles or rotating vortices. Their rotation shifted small cog wheels between the vortices; the displacement of the cog wheels represented electric current. Lines of $D$ or $E$ paralleled these currents. This was Maxwell's model of the displacement current in vacuum; there was vigorous debate in the 1880's as to the real nature of this current.

(Except $\boldsymbol{j}_{mag} = c\nabla \times \boldsymbol{M}$ in Gaussian and Heaviside-Lorentz units). The total charge density is

$$\rho = \rho_{free} + \rho_{pol} = \rho_{free} - \nabla \cdot \boldsymbol{P} \tag{24}$$

where $\rho_{free}$ is the charge density of unbound (free) charges. The total current density is

$$\boldsymbol{j} = \boldsymbol{j}_{free} + \boldsymbol{j}_{pol} + \boldsymbol{j}_{mag} = \boldsymbol{j}_{free} + \frac{\partial \boldsymbol{P}}{\partial t} + \nabla \times \boldsymbol{M} \tag{25}$$

where $\boldsymbol{j}_{free}$ is the current density of freely moving charges, including conduction current $\sigma \boldsymbol{E}$. (Since $\nabla \cdot \boldsymbol{j}_{mag} = 0$, there is no *electric* charge density associated with magnetization [Gr13], but $\nabla \cdot \boldsymbol{M}$ can be considered a *magnetic charge* density [Ja75].)

$\boldsymbol{P}$ and $\boldsymbol{M}$ are sources of fields, and as such must be included in Maxwell's Equations. The advantage of defining $\boldsymbol{D}$ and $\boldsymbol{H}$ is: when $\boldsymbol{D}$ instead of $\boldsymbol{E}$ is used in Gauss' Law, the law remains in simple form, the induced polarization being taken care of "automatically", leaving only the free charge density on the right hand side. Similarly, when $\boldsymbol{H}$ and $\boldsymbol{D}$ are used instead of $\boldsymbol{B}$ and $\boldsymbol{E}$ in Ampere's Law, the equation retains its simplicity; magnetization currents and polarization currents are included automatically, leaving only currents $\boldsymbol{j}_{free}$ due to freely moving particles on the right hand side. That is, they incorporate the effects of matter in a very simple way. $\boldsymbol{E}$ is driven by the total charge density; $\boldsymbol{D}$ is driven by free charges[43]. $\boldsymbol{B}$ is driven by the total current; $\boldsymbol{H}$ is driven by free currents.

The two equations (21) appear oddly unsymmetrical. Paraphrasing Cohen [Co01]:

> In the 19$^{th}$ century, $\boldsymbol{E}$, the electric field intensity, and a point charge $q$, were considered the fundamental electrical quantities; the field of a point charge in vacuum was written $\boldsymbol{E} = q/\epsilon_o r^2$, with $\epsilon_o$ a generic unit-dependent parameter. $\boldsymbol{D}$ was considered a useful auxiliary field in matter, as it is today. In free space one would have written $\boldsymbol{D} = \epsilon_o \boldsymbol{E}$ as we do now.
>
> Physicists at that time were probably less certain than today that a magnetic monopole does not exist, although poles always seemed to occur in pairs. It was common to consider the magnetic field of a monopole $p$, as Gauss did in many of his theoretical developments. As with $\boldsymbol{E}$, the magnetic field $\boldsymbol{H}$, comparably called the magnetic field intensity, and its associated magnetic poles were regarded as the basic theoretical ideas; $\boldsymbol{H}$ of a point monopole was written $\boldsymbol{H} = p/\mu_o r^2$, with $\mu_o$ a generic unit-dependent parameter. $\boldsymbol{B}$ was treated as more of a useful auxiliary construct for fields in matter, so, similar to $\boldsymbol{D} = \epsilon_o \boldsymbol{E}$, one would have written $\boldsymbol{B} = \mu_o \boldsymbol{H}$ in free space. Later in the 19$^{th}$ century, and into the 20$^{th}$, understanding about the origin of material magnetization improved, Faraday's law of induction (involving $\boldsymbol{B}$) was better understood, and the recognition of the Lorentz force law, involving **B**, not **H**, caused these views of **H** and **B** to be reversed; **B** was now considered the fundamental field. Thus the usual way of writing switched to $\mathbf{H} = (1/\mu_o)\mathbf{B}$.

---

[43] For a caveat in solving Gauss' Law $\nabla \cdot \boldsymbol{D} \propto \rho_{free}$, see Griffiths [Gr13], Section 4.3.2.

This explains why, in relations (21), **E** is multiplied by the parameter $\epsilon_o$, but **B** is divided by $\mu_o$, and why **M**, which would have appeared as $\mathbf{B} = \mu_o\mathbf{H} + \lambda''\mathbf{M}$, has the minus sign.

Dimensions need to be chosen for **D** and **H** and the four new constants. **P** already has dimensions $Q \cdot L/L^3 = Q/L^2$, and **M** is $I \cdot L^2/L^3 = I/L$. Jackson argues that nothing is gained by making different dimensions for **D** and **P**, or for **H** and **M** so $\lambda$ and $\lambda'$ are taken to be dimensionless numbers ($\lambda = \lambda' = 1$ in rationalized systems; $\lambda = \lambda' = 4\pi$ in unrationalized systems).

The values chosen in the various systems for $\epsilon_o$ and $\mu_o$ in Eq (21) are shown in Table 2, taken from Jackson's appendix.

**Table 2. Values of $\epsilon_o$ and $\mu_o$ in several systems [Ja75]**

|  | esu | emu | Gaussian | Variant Gaussian | H-L | SI |
|---|---|---|---|---|---|---|
| $\epsilon_o$ | 1 | $1/c^2$ (s$^2$/cm$^2$) | 1 | 1 | 1 | $1/\mu_o c^2 \approx 8.854 \times 10^{-12}$ (F/m) |
| $\mu_o$ | $1/c^2$ (s$^2$/cm$^2$) | 1 | 1 | 1 | 1 | $4\pi \times 10^{-7}$ (H/m) |

Although in (21) $\epsilon_o$ is a "new constant" to be defined, the notation belies forethought in SI units; only in SI is the symbol $\epsilon_o$ retained as a displayed constant. In that system it may seem curious that the same $\epsilon_o$ that appears in the original Coulomb force law $q^2/4\pi\epsilon_o r^2$ ($k_e = 1/4\pi\epsilon_o$) should also occur in the relation (21) between the displacement vector **D** and the electric field **E**, involving matter and so not especially related to the vacuum force law. One thinks it ought to be a new, separate constant. The reason is as follows. In SI $\epsilon_o$ has its conventional definition $\epsilon_o = 1/\mu_o c^2 \approx 8.85 \times 10^{-12}$ F/m. And Ampere's Law in (20) is put in its conventional form in terms of **H** by dividing by $\mu_o$, so the coefficient of $\partial \mathbf{E}/\partial t$ becomes $k_B/\mu_o c^2 = \epsilon_o$, and the term in $\partial \mathbf{E}/\partial t$ is the usual displacement current $\partial \mathbf{D}/\partial t$. So the same $\epsilon_o$ in the force law does correctly appear multiplying **E** in the constitutive relations.

Although $\mu_o$ is commonly called the permeability of the vacuum, it is coming to be called ***the magnetic constant***, emphasizing that it is a definition, with value that varies from one unit system to another, and deemphasizing any relation to a physical property of a medium, even a vacuum.

For the same reason $\epsilon_o$ is coming to be called ***the electric constant***.

Because of the constitutive relations (21) Maxwell's Equations in arbitrary units, including material media, would involve $k_e$, $k_B$, $\epsilon_o$, $\mu_o$, $\lambda$, and $\lambda'$. Little is gained by displaying them in such generality. They will be presented below in the individual systems.

We proceed to discuss the systems that are important today. Although esu and emu are obsolete, they are the building blocks of the widely used *Gaussian system,* and must be discussed first.

# 6 THE VARIOUS SYSTEMS

## 6.1 THE <u>ELECTROMAGNETIC</u> UNITS SYSTEM

We begin with the ***Electromagnetic Units (emu)*** system, as it was historically the first to be developed. Pure emu and pure esu have not been used for many decades. Both have been replaced by the Gaussian system, a mix of the two. But to understand Gaussian units it is really necessary to understand emu and esu separately.

In emu only mass, length, and time occur; no new dimensional quantity is employed. The system (but not its name) was introduced by Gauss in the 1830's for his measurements of the geomagnetic field. Prior to his work, physicists could only report that a measured force or field was some factor larger or smaller than another; there were only measures of one relative to another.

Gauss sought an absolute measure, and argued that since it was difficult to maintain physical laboratory magnetic standards that hold their values over long times, and that are reproducible worldwide, it was better to define magnetic quantities directly in terms of the mechanical effects they produce (yet see footnote 29). Hence he retained only gram, cm, and sec[44], and found a way to measure magnetic fields entirely in terms of those three base units. Hence, emu is a cgs system with no other base dimension. Gauss' method is summarized in Appendix A.

As it was introduced for (static) geomagnetism, why is it not called the *magnetostatic system*? Because at the time (1830's) Wilhelm Weber was working with Gauss, exploring Oersted's electric-magnetic connections of 1820. He found Gauss' new method could be extended to his electrical quantities. The system became widely used, and sanctified by the 1861 B.A. committee, only after it had been employed in electromagnetic phenomena, and there was no need to restrict its name to static conditions[45, 46].

The Gauss-Weber units was the world's first quantitative method to measure electric and magnetic quantities in terms of known mechanical units. Into the 1850's their system was developed and promoted, under the active leadership of Maxwell and William Thomson (Lord Kelvin) (1824-1907), through the British Association for the Advancement of Science. As mentioned, the 1862 BA committee assigned the name "Electromagnetic Units System (emu)" to the Gauss-Weber system, which previously had no particular name.

---

[44] Oddly, in their own work Gauss and Weber used the millimeter, milligram, and second as base units [Si62; So52, p.42].

[45] If a magnetic monopole existed, the force between two of them would be much greater than the electrostatic force between two electrons, and magnetostatics as a subject in its own right would no doubt have been at the forefront of research long before it was. An early unit system for it then might well be known as the magnetostatic system.

[46] Parenthetically, it is curious that in 1833 Gauss and Weber created the world's first telegraph, a line between Gottingen University and its observatory some 3 km away. Some scholars credit Sir Charles Wheatstone (1802-75), of bridge fame (and uncle of Oliver Heaviside), and William F. Cooke as the inventors of the telegraph. Still others credit Samuel F.B. Morse and Joseph Henry. It is likely that the Gauss-Weber device predates the others by a year or two.

In textbook discussions of emu, it is stated that, in contrast to electrostatic units, current, rather than charge, is chosen as the quantity to connect to mechanics[47]. That is true, and we will continue with the development starting with Eq (4). But historically Gauss did not start with Eq (4). Rather, as was common at the time, Gauss started with the magnetic analog of the electrostatic Coulomb force, Eq (3); namely

$$f_{\text{magnetostatic}} = k \frac{p_1 p_2}{r^2} \tag{26}$$

for the force between two magnetic poles $p_1$ and $p_2$. For simplicity the constant $k$ was chosen to be 1, dimensionless. Thus historically the electromagnetic system of units was based on the unit magnetic pole $p = \text{dyn}^{1/2} \cdot \text{cm}$. It was recognized that no magnetic monopole exists, or at least that none had ever been seen. It was understood then that the ends of a long, thin rod magnet approximate poles; the $B$ field of a single pole drops off as $1/r^2$, just as the electrostatic field of a point charge. Although Gauss' theoretical approach started with (26), his experiments were, of course, with magnetic dipoles, lodestones machined to a convenient shape, or magnetized steel needles.

Historically the method based on (26) gave way to that based on (4) to eliminate reliance on a fictional pole, and to base the development on a realizable current and force. Thus we continue with Eq (4), the familiar method today. In emu, the unit of current is chosen so that Ampere's magnetostatic force law is simple: the force (4) between two fine, equal line currents $I$ separated by $r$ is set equal to

$$f' = 2\frac{I^2}{r} \equiv f'_{\text{emu}} \tag{27}$$

That is, referring to the basic Eq (4), emu chooses $k_m = 1$, so the overall proportionality constant is defined[34] as 2, a dimensionless pure number. The unit of **current** is then defined as that current which causes a force of $f' = 2$ dyn/cm at $r = 1$ cm. The name given to this unit by the 1862 BA committee is 1 *abAmpere* (abA)[48, 49]. The dimensions of current are then force$^{1/2}$. Thus, one abA automatically has dimension and magnitude

$$1 \text{ abA} = 1 \text{ emu unit of current} = 1 \text{ dyn}^{1/2}, \tag{28}$$

---

[47] Presumably because laboratory magnetic studies used currents to create and control the field, even if it was difficult to maintain them for long times. Electric charges were difficult to maintain.

[48] The "ab" comes from "absolute". The name "absolute" system of electromagnetic units was used historically to mean two things: 1) before Weber's and Gauss' work measurements of charges or currents were only relative to another charge or current. They were just comparisons between magnitudes of the same kind. No absolute measure existed until Gauss' new system [We91]. Gauss' paper was entitled *The intensity of the earth's magnetic force reduced to absolute measurement*. An English translation by S P Johnson can be found at [Jo95]. Measures were made absolute by expressing them in terms of cm, g, and sec. In these uses of the term, both esu and emu were absolute systems. The word means no more than that; there is nothing fundamentally absolute about emu or any other unit systems. And 2) importantly, absolute units do not rely on physical standards (except those of mass, length, and time), unlike units in the "International System". The prefixes "ab" and "abstat", later shortened to "stat", were apparently proposed by Kennelly in 1903 [Ke03; Cu37, p. 11].

[49] The abAmpere was at first called the Biot by the committee.

with dimensions $M^{1/2} L^{1/2} T^{-1}$. Relative to the other systems it will turn out that 1 abampere = $3 \times 10^{10}$ electrostatic units of current (stA)[50] = 10 Amperes[51].

In electromagnetic units, the unit of **charge**, the *abCoulomb* (abC) follows directly from (5), and is simply

$$1 \text{ abC} = 1 \text{ emu unit of charge} = 1 \text{ abA} \cdot 1 \text{sec} = 1 \text{ dyn}^{1/2} \cdot \text{sec} \qquad (29)$$

with dimensions $M^{1/2} L^{1/2}$. Relative to SI, 1 abC = 10 Coulomb. The charge of an electron is $1.602 \times 10^{-20}$ abC = $1.602 \times 10^{-19}$ C.

> At first view we have an unsettling feeling. Intuitively a $dyn^{1/2}$ has nothing to do with a current, and $dyn^{1/2}$·sec is not very charge-like. We feel better with a separate electromagnetic, non-mechanical unit. But many physicists have emphasized that the dimensions and units of any physical quantity need not bear any relation to intuitive expectations. Natural units, in which many quantities are dimensionless, drive this point home. Even in SI, in which we are all comfortable with the Ampere and Coulomb, any comfort is an illusion. The SI dimension of current avoids odd things like $dyn^{1/2}$ only because of definitions, and an Ampere is still defined in terms of force. Comfort is mere familiarity after long use following one's first introduction to the subject.

With $k_m = 1$, $k_e$ must be $c^2$ (Eq (7)), and the force between two point charges (abC) is[52]

$$f = c^2 \frac{q_1 q_2}{r^2} = f_{emu} \qquad (30)$$

The force between two electrons 1 cm apart is then $(3 \times 10^{10})^2 \times (1.602 \times 10^{-20})^2/(1 \text{ cm})^2 = 2.31 \times 10^{-19}$ dyn, which, of course, will agree with all systems. Likewise, the potential energy of two charges separated by $r$ is $c^2 q_1 q_2 / r$ (erg). Gauss' work preceded Heaviside. There is no $4\pi$ in the denominator of $f$; the emu system is unrationalized.

In emu the unit of **potential** is the *abVolt*,

$$1 \text{ abV} = 1 \text{ emu unit of potential} = 1 \text{erg}/1 \text{abC} = \text{dyn}^{1/2} \cdot \text{cm/sec} \qquad (31)$$

with dimensions $M^{1/2} L^{3/2} T^{-2}$. Relative to SI, a potential of 1 abV is equal to $1 \times 10^{-8}$ Volt.

The unit of **electric field** in emu is

$$E_{emu} = 1 \text{ emu unit of electric field} = 1 \text{dyn}/1 \text{abC} = 1 \text{abV/cm} = \text{dyn}^{1/2}/\text{sec} \qquad (32)$$

with dimensions $M^{1/2} L^{1/2} T^{-2}$, and has no separate name. Relative to SI, a field of 1 abV/cm (1 emu of electric field) is $10^{-6}$ V/m.

As from (11), the dimensions of ***B*** are

$$[\boldsymbol{B}] = [k_B][\boldsymbol{E}/c].$$

---

[50] The factor 3 is actually 2.99792... from the speed of light.
[51] Historically the abAmpere came first. When the Ampere was defined, it was arbitrarily (yet with good reason) set to be $10^{-1}$ abAmpere. *cf* Section 6.7.
[52] The abCoulomb is not used in any other system and has fallen completely out of use.

In emu the **magnetic field** (10) is chosen to have the simplest relation to $I$:

$$B = 2\frac{I}{r} \quad \text{(emu)} \tag{33}$$

that is, the choice made is $k_B = 1$. $I$ is in abA. Dimensionally, $[B] = [E/c]$. This gives $B$ the dimension abA/cm, or $M^{1/2} L^{-1/2} T^{-1}$, the same as $[E/c] = $ abV·s/cm$^2$ = abA/cm.

In emu the **continuity** law of charge and current reads the conventional way, Eq(8)

$$\nabla \cdot \boldsymbol{j} + \frac{\partial \rho}{\partial t} = 0 \tag{34}$$

with $\boldsymbol{j}$ in abA/cm$^2$, and $\rho$ in abC/cm$^3$.

### 6.1.1 THE UNIT OF MAGNETIC FLUX DENSITY $B$

The unit for $B$ is defined in terms of magnetic flux. The unit of magnetic flux is called the **maxwell** (Mx)[53], and is defined with reference to Faraday's Law in integral form,

$$\text{emf} = \oint \boldsymbol{E} \cdot d\boldsymbol{l} = -\frac{d\phi}{dt} \tag{35}$$

where $\phi$ is the magnetic flux through an area bounded by the closed loop of the integral. The maxwell is defined as

$$\begin{aligned} &1 \text{ maxwell} \equiv 1 \text{ Mx} \equiv \text{emu unit of magnetic flux} = \text{that flux, changing} \\ &\text{at rate 1 Mx/sec, that induces an emf of 1 abV, or} \\ &1 \text{ \textbf{maxwell}} = 1 \text{ abV} \cdot \text{sec} = 1 \text{ dyn}^{1/2} \cdot \text{cm} \end{aligned} \tag{36}$$

Relative to SI, 1 maxwell = $10^{-8}$ weber $\equiv 10^{-8}$ Wb[54].

The unit of magnetic flux density $B$ is defined as

$$\begin{aligned} \textbf{1 gauss} &\equiv \text{the emu unit of magnetic flux density} \\ &= 1 \text{ Mx/cm}^2 = 1 \text{ abA/cm} = 1 \text{ dyn}^{1/2}/\text{cm}. \end{aligned} \tag{37}$$

Relative to SI, 1 gauss = $10^{-4}$ Tesla.

From this definition, one deduces:

- Since a field $B$ exerts a force $f' = IB$, or $f'(\text{dyn/cm}) = I(\text{abA})B(\text{abA/cm})$, on a current $I$, 1 gauss is that field which exerts a force of 1 dyn/cm on a perpendicular wire carrying 1 abA.

- 1 gauss is that field which induces an emf of 1 abV/cm along a wire moving at 1 cm/s through it.

---

[53] In the late 19$^{th}$ century the maxwell was also called the *line*, in reference to graphical depiction of field lines. The closer lines were together, the higher the flux density and the more intense the field.

[54] SI nomenclature etiquette specifies that a unit name, when spelled out, shall be all lower case. When abbreviated by its symbol, the initial letter is to be capitalized, 1 weber = 1 Wb, 1 pascal = 1 Pa, 1 tesla = 1 T. The name shall furthermore always be in roman, non-italic type(!) [BI06]

- 1 gauss is the field 1 cm from 0.5 abA in a fine wire, Eq (33).

These statements are occasionally taken as equivalent definitions of 1 gauss[55].

### 6.1.2 THE UNIT OF MAGNETIC FIELD *H*

Into the early 20th century there was much confusion between the fields ***B*** and ***H***, and before 1930 there was similar confusion about the difference between the gauss (the emu and Gaussian unit of magnetic flux density ***B***) and the oersted (the unit of ***H***). At its meeting in Stockholm in 1930 the Advisory Committee on Nomenclature of the International Electrotechnical Commission eliminated all ambiguity by adopting the gauss for the unit of magnetic flux density and the oersted for the unit of magnetic field strength. The gauss is defined through a time-dependent change in magnetic flux density and the oersted is defined through the field created by an electric current[56]. That definition is as follows.

In steady state, the integral form of Ampere's Law from (48) is

$$\text{mmf} = \oint \bm{H} \cdot \bm{dl} = 4\pi \int \bm{j} \cdot \bm{dS} = 4\pi I \tag{38}$$

defining the *magnetomotive force mmf* around the circuit of the line integral. Solving for ***H*** a distance *r* from a wire current *I*,

$$H = 2\frac{I}{r} \qquad \text{(emu)} \tag{39}$$

the same as *B*. Both ***B*** and ***H*** have the dimensions of a current per unit length. The *unit* of *H* is called the oersted, defined by

$$\begin{aligned}\textbf{1 oersted} &\equiv \text{the emu unit of magnetic field strength } H \\ &= \text{the field 1 cm from a wire carrying 0.5 abA} \\ &= 1 \text{ abA/cm.}\end{aligned} \tag{40}$$

the same as *B*. Relative to SI, it will turn out that a field of 1 oersted is the same as a field of $1000/4\pi$ A/m. See Appendix B regarding this and other "equivalences" between systems, or "conversion factors", and common pitfalls in using them.

### 6.1.3 CONSTITUTIVE RELATIONS

From Eq (21) and Table 2, in emu one has ***D*** and ***H*** defined as

$$\begin{aligned}\bm{D} &= \frac{1}{c^2}\bm{E} + 4\pi\bm{P} \\ \bm{H} &= \bm{B} - 4\pi\bm{M}\end{aligned} \tag{41}$$

---

[55] Curiously, some textbooks state the unit of magnetic field (e.g. the gauss or tesla), but do not specifically define it. From equations such as, for example, (33), $B(\text{gauss}) = 2I(\text{abA})/r(\text{cm})$, the reader is left to infer that a realization (or definition) of 1 gauss is the field 1 cm from a wire carrying 0.5 abA.

[56] http://en.citizendium.org/wiki/Oersted_(unit)

with dimensions $[D] = [P] = \text{abC/cm}^2$, $[E] = \text{abV/cm} = \text{abC/s}^2$; $[H] = \text{oersted} = [B] = [M] = \text{gauss} = \text{abA/cm}$.

### 6.1.4 OTHER DYNAMICAL QUANTITIES

According to (12), the **Lorentz force in emu** reads

$$F_L = q(E + v \times B) \ . \tag{42}$$

or, with units spelled out:

$$F_L(\text{dyn}) = q(\text{abC})\big(E(\text{abV/cm}) + v(\text{cm/s}) \times B(\text{gauss})\big) \ . \tag{43}$$

From this one obtains the **force per unit length** $f'$ on a wire carrying current $I$ in an external field $B$

$$f'(\text{dyn/cm}) = IB \ , \tag{44}$$

with $I$ in abA and $B$ in gauss.

The relation between the magnitudes $E$ and $B$ fields in a **plane wave in vacuum** is obtained by inserting space and time dependence $\exp(ikx - i\omega t)$, with $k = \omega/c$, in Faraday's Law, $ik \times E - i\omega B = 0$, to obtain $E = cB$, meaning

$$E(\text{abV/cm}) = c(\text{cm/s})B(\text{gauss}) = 3 \times 10^{10} B(\text{gauss}) \ .$$

The **resistance unit** is

$$1 \text{ abOhm} \equiv 1 \text{ abV}/1 \text{ abA} = 1 \text{ cm/sec},$$

with the dimensions of velocity. Relative to SI, $1 \text{ ab}\Omega$ corresponds to $10^{-9} \ \Omega$.

The **Impedance of Free Space** $Z_o$ is the ratio $E/H$ in a plane wave. Since $H(\text{Oe}) = B(\text{gauss})$, $Z_o$ is

$$Z_o = E(\text{abV/cm}) / H(\text{Oe}) = 3 \times 10^{10} \text{ cm/s} = 3 \times 10^{10} \text{ ab}\Omega = c.$$

The resistance of a wire of length $L$ and cross sectional area $A$ is $R = \rho L/A$, where $\rho$ is the material resistivity. Thus resistivity has dimension $\text{ab}\Omega \cdot \text{cm} = \text{cm}^2/\text{sec}$, and conductivity, $\sigma = 1/\rho$, $\text{sec/cm}^2$.

The **capacitance unit** is

$$1 \text{ abFarad} \equiv 1 \text{ abC}/1 \text{ abV} = 1 \text{ sec}^2/\text{cm},$$

with the dimensions of 1/acceleration. Relative to SI, $1 \text{ abFarad} = 10^9$ Farad.

**Inductance** $L$ is defined by (magnetic flux) = $L \cdot$(Current), so the unit of inductance is

$$1 \text{ abHenry} = 1 \text{ maxwell}/1 \text{ abA} = 1 \text{ gauss} \cdot \text{cm}^2/\text{abA} = 1 \text{ cm}$$

with dimensions of length. Relative to SI, $1 \text{ abHenry} = 10^{-9}$ Henry. The induced emf is

$$\text{emf}(\text{abV}) = L(\text{cm}) \, dI(\text{abA})/dt.$$

**Energy flow** in the fields is obtained in the usual way of writing the identity
$\nabla\cdot(\boldsymbol{E}\times\boldsymbol{H}) = \boldsymbol{H}\cdot\nabla\times\boldsymbol{E} - \boldsymbol{E}\cdot\nabla\times\boldsymbol{H}$ and replacing the curls from Faraday's and Ampere's Laws to obtain the **Poynting vector**

$$\boldsymbol{S} = \frac{1}{4\pi}\boldsymbol{E}\times\boldsymbol{H}: \qquad \boldsymbol{S}(\text{erg/cm}^2/\text{sec}) = \frac{1}{4\pi}\boldsymbol{E}(\text{abV/cm})\times\boldsymbol{H}(\text{oersted}) \qquad (45)$$

and the **energy density**

$$u = \frac{1}{8\pi}(\boldsymbol{B}\cdot\boldsymbol{H} + \boldsymbol{E}\cdot\boldsymbol{D}) \quad (\text{erg/cm}^3),$$

giving **energy conservation**

$$\nabla\cdot\boldsymbol{S} + \frac{\partial u}{\partial t} = -\boldsymbol{j}\cdot\boldsymbol{E} \qquad (46)$$

The $1/4\pi$ in defining $\boldsymbol{S}$ and $u$ allow the rhs of (46) to be the standard Joule heating term $\boldsymbol{j}\cdot\boldsymbol{E}$ = (abA/cm$^2$)·(abV/cm) = abA·abV/cm$^3$ = erg/cm$^3$/sec. In a plane wave $E = cB$, and $S = cu$.

Above we have presented many quantities. Continuing in this manner, in emu and in all systems, one develops expressions for the units of all dynamical quantities of interest. We will present fewer in later systems.

**The Fine Structure Constant**
In general the fine structure constant $\alpha$ is the dimensionless ratio of the electrostatic energy of two electrons separated by a Compton wavelength ($\lambda_C = \hbar/mc$), to the electron rest energy $mc^2$. Because of (30), in emu it is written $(c^2e^2/\lambda_C)/mc^2$, or

$$\alpha = c^2e^2/\hbar c = c(\text{cm/s})^2 e(\text{abC})^2/\hbar(\text{erg·s})c(\text{cm/s}) = 1/137 \qquad (\text{emu}).$$

Being dimensionless it has the same numerical value in all systems, but due to different units for charge its expression in terms of $e$, $\hbar$, and $c$ differs in different unit systems. As discussed momentarily, in electrostatic units and in Gaussian units, it appears written as $\alpha = e^2/\hbar c$, and in SI it is $\alpha = e^2/4\pi\epsilon_o\hbar c$.

### 6.1.5  MAXWELL'S EQUATIONS IN ELECTROMAGNETIC UNITS
According to Eq (20), in emu, with $k_m = k_B = 1$, $k_e = c^2$, $k_F = 1$, $\epsilon_o = c^{-2}$, $\mu_o = 1$, Maxwell's Equations in the absence of matter ($\boldsymbol{P} = \boldsymbol{M} = 0$) are

$$\begin{aligned}\nabla\cdot\boldsymbol{E} &= 4\pi c^2\rho \\ \nabla\times\boldsymbol{B} - \frac{\partial\boldsymbol{E}}{c^2\partial t} &= 4\pi\boldsymbol{j} \\ \nabla\times\boldsymbol{E} + \frac{\partial\boldsymbol{B}}{\partial t} &= 0 \\ \nabla\cdot\boldsymbol{B} &= 0\end{aligned} \qquad (47)$$

displaying powers of $c$ in certain terms. Incorporating matter with the constitutive relations (41), these become

$$\nabla \cdot \boldsymbol{D} = 4\pi\rho$$
$$\nabla \times \boldsymbol{H} - \frac{\partial \boldsymbol{D}}{\partial t} = 4\pi \boldsymbol{j}$$
$$\nabla \times \boldsymbol{E} + \frac{\partial \boldsymbol{B}}{\partial t} = 0 \qquad (48)$$
$$\nabla \cdot \boldsymbol{B} = 0$$

The dimensions are $\rho$(abC/cm$^3$), $\boldsymbol{j}$(abA/cm$^2$), $\boldsymbol{E}$(abV/cm), $\boldsymbol{D}$(abC/cm$^2$), $\boldsymbol{H}$(abA/cm = oersted), and $\boldsymbol{B}$(abA/cm = gauss).

Historically, strict emu was used for most of the early history[57] of electromagnetism; at least one book was written entirely in emu [Gi32]. But its units of practical, laboratory quantities were inconvenient. As stated, the unit of emf, the abVolt, is 1 abV = $10^{-8}$ Volt, inconveniently small. Likewise, its unit of resistance, 1 abOhm = 1 abVolt/abAmp, is $10^{-9}$ Ohm. Hence, like esu, pure emu is essentially never used today[58].

One can compare the matter-free Maxwell's Equations in emu, (47), with the same equations in other units to be discussed and watch factors of $c$ flit from term to term.

## 6.2 THE ELECTROSTATIC UNITS SYSTEM

Although the electromagnetic system was clearly invented by Gauss and Weber, the official ***Electrostatic Units (esu)*** system snuck into existence as a by-product of a committee's observations.

This systems starts with the electrostatic force in simplest form

$$f = \frac{q_1 q_2}{r^2} \equiv f_{esu} \ . \qquad (49)$$

That is, $\boldsymbol{k_e}$ is chosen to be **1**, and one does not introduce any new dimension for charge.

The *unit* of **charge** is set by defining one unit as that amount of charge which causes a force of 1 dyne when separated by $r = 1$ cm from another equal charge, $f = q^2/(1\,\text{cm})^2 = 1$ dyn. The name given to this unit is 1 ***statCoulomb***, stC (or, colloquially, one *esu*)[59]. Then 1 dyn = $(1\,\text{stC})^2/(1\,\text{cm})^2$, and the dimensions of charge are $[q] = [\text{stC}] = \text{dyn}^{1/2}\cdot\text{cm}$, or $[q^2]$ = erg·cm ~ M·L$^3$/T$^2$. Thus, one stC automatically has dimension and magnitude

$$1\,\text{stC} = 1\,\text{esu unit of charge} = 1\,\text{dyn}^{1/2}\cdot\text{cm} \sim M^{1/2} L^{3/2} T^{-1} \qquad (50)$$

---

[57] Ampere himself used another system called *Electrodynamic Units*. It was later abandoned. See [As03] and [Ma91, Art. 526].

[58] One could say that in SI the unit of charge, the Coulomb, is inconveniently big, at least to express the charge of an electron. But for practical circuit work, 1 Amp is (was) a practical current unit, and 1 second is a reasonable time unit. So 1 C ≡ 1 A×1 sec is stuck with its value, and is convenient for laboratory work, at least in those uncommon circumstances in which one wants to know the amount of charge passing in a wire. But typical circuit capacitors (pF to μF) hold charges measured in pC to μC.

[59] The term *esuch* was once suggested as more appropriate than statCoulomb [Sh39]. The electric field unit would be 1 dyn per esuch. The statCoulomb is also occasionally called the ***Franklin***.

Relative to other systems, 1 Coulomb = $3 \times 10^9$ stC, and 1 abC = $3 \times 10^{10}$ stC [60]. The charge of an electron is $4.803 \times 10^{-10}$ stC, and the force between two electrons separated by 1 cm is

$$f = (4.803 \times 10^{-10})^2/(1\,\text{cm})^2 = 2.31 \times 10^{-19} \text{ dyn.}$$

as in emu of course.

It is clear from (29) and (50) that the *dimensions* of charge in emu ($\text{dyn}^{1/2}\cdot\text{sec}$) are different from those in esu ($\text{dyn}^{1/2}\cdot\text{cm}$). This property, that a physical quantity can have different dimensions in different units systems, is a feature that occurs in electromagnetism that does not occur in mechanical units. In the latter, all units of length, say, have the same dimensions; so the conversion factor from one unit system to another is a pure dimensionless number, like the 2.54 in 1 inch = 2.54 cm. But in electromagnetism the freedom of starting with either (4) [or (27)] or with (3) [or (49)] allows different dimensions for charge, current, and other electromagnetic quantities. So the conversion factor from one unit system to another itself has dimensions. This circumstance can cause much confusion. Appendix B discusses common pitfalls when comparing charges and other quantities in different unit systems.

In esu the unit of **current** is charge/sec, called the *statAmpere* (stA) [61],

$$1\,\text{stA} = 1 \text{ esu unit of current} = 1\,\text{stC/sec} = 1\,\text{dyn}^{1/2}\cdot\text{cm/sec} \sim M^{1/2}L^{3/2}T^{-2} \qquad (51)$$

1 stA is a very small current. Relative to emu, 1 stA is $1/(3 \times 10^{10})$ abA = $3.33 \times 10^{-11}$ abA, and relative to SI, 1 stA is $1/(3 \times 10^9)$ A = $3.33 \times 10^{-10}$ Ampere. Having chosen $k_e = 1$, we must have $k_m = 1/c^2$, so that in esu Eq (4), the force between two currents necessarily is

$$f' = \frac{2}{c^2}\frac{I_1 I_2}{r} = f'_{\text{esu}} \qquad (52)$$

Here, of course, $c = 3 \times 10^{10}$ cm/s. The $I$'s are in stA.

In esu the unit of **potential**, *statVolt*, can be defined by

$$1\,\text{stV} = 1 \text{ esu unit of potential} = 1\,\text{erg}/1\,\text{stC} = 1\,\text{dyn}^{1/2} \sim M^{1/2}L^{1/2}T^{-1} \qquad (53)$$

It turns out that a potential of 1 stV is equal to 300 Volts in SI.

The **electric field** unit in esu is

$$E_{\text{esu}} = \text{esu unit of electric field} = 1\,\text{dyn}/1\,\text{stC} = 1\,\text{stV/cm} = 1\,\text{dyn}^{1/2}/\text{cm} \sim M^{1/2}L^{-1/2}T^{-1} \qquad (54)$$

It has no separate name. A field of 1 stV/cm (or "1 esu" of electric field) is $3 \times 10^4$ V/m.

---

[60] Instead of discussing the ratio of constants $\sqrt{(k_e/k_m)}$, Eq(7), Maxwell put it: "The number of electrostatic units of electricity contained in one electromagnetic unit is numerically equal to a certain velocity … . This velocity is an important physical quantity, which we shall denote by the symbol $\upsilon$" [Ma91, Art. 628]. Physicists later changed the symbol to $c$.

[61] One wonders why a separate name is needed for a unit of such a simple thing as charge/time. There is, for example, no separate name for a unit of electric field; it is simply V/m. No one has ever bothered to assign a name for a unit of length/time, and no one objects to saying "m/s" for velocity. For the speed of sound, notwithstanding the Mach number, we are quite content with "341 m/s" rather than, say, "1 Yeager".

As discussed later, the factor $4\pi$ that appears in the electrostatic Coulomb force law (3) in SI units ($q_1q_2/4\pi\epsilon_o r^2$) does not appear in that law expressed in esu; rather it occurs explicitly in Maxwell's Equations, making esu, like emu, an *unrationalized* system (*cf* Section 6.5). The esu system was set up before Heaviside's rationalization urgings.

In esu the continuity law reads the conventional way, Eq (34),

$$\nabla \cdot \boldsymbol{j} + \frac{\partial \rho}{\partial t} = 0 \tag{55}$$

but now with $\boldsymbol{j}$ in stA/cm$^2$, and $\rho$ in stC/cm$^3$.

The **magnetic field** at $r$ from a wire is obtained from (10). In esu the choice made is $k_B = 1$, dimensionless, so that, with $k_m = 1/c^2$, the dimensions of $\boldsymbol{B}$ are $[\boldsymbol{B}] = [I/c^2 r] = [Q/cr^2] = [\boldsymbol{E}/c] = [\text{stV·s/cm}^2] = M^{1/2}L^{-3/2}$. The field from a current is then:

$$B = \frac{2}{c^2}\frac{I}{r} \qquad \text{(esu)} \tag{56}$$

This unit appears to have no particular name, except perhaps the construct *statgauss*. The magnetic flux unit might be called the *statmaxwell* = 1 statgauss·cm$^2$, with dimensions stV·sec. Magnetic quantities in esu are not very transparent. As esu has not been used for magnetic quantities for a century, there is really no occasion to employ Eq (56) today.

### 6.2.1 MAXWELL'S EQUATIONS IN ELECTROSTATIC UNITS

The choices $k_e = 1 = k_B$, leave $k_m = 1/c^2$, $k_F = 1$, so that Maxwell's Equations in esu, in the absence of material media, are

$$\begin{aligned} \nabla \cdot \boldsymbol{E} &= 4\pi\rho \\ \nabla \times \boldsymbol{B} - \frac{1}{c^2}\frac{\partial \boldsymbol{E}}{\partial t} &= 4\pi\frac{1}{c^2}\boldsymbol{j} \\ \nabla \times \boldsymbol{E} + \frac{\partial \boldsymbol{B}}{\partial t} &= 0 \\ \nabla \cdot \boldsymbol{B} &= 0 \end{aligned} \tag{57}$$

The units are $\rho$(stC/cm$^3$), $\boldsymbol{j}$(stA/cm$^2$), $\boldsymbol{E}$(stV/cm=stC/cm$^2$), and $\boldsymbol{B}$(stV·s/cm$^2$).

Relative to emu, Eqs (47), the choice of parameters in esu effectively interchanges the roles of $k_e$ and $k_m$, moving the factor $c^2$ on the right hand sides of Maxwell's Equations from the $\rho$ term to $\boldsymbol{j}$.

Except for Coulomb's Law in the form (49), Electrostatic Units have not been widely used, if at all, for more than 100 years, and almost certainly never will be used again. Except in textbook appendices on units, likely no one has bothered to write down Eqs (57) for a century. The reader might as well forget them immediately. But other esu equations, especially (49), do form part of the basis for the widely used Gaussian system.

Some other quantities are:

The **Lorentz force in esu** reads

$$\boldsymbol{F}_\mathrm{L}(\mathrm{dyn}) = q(\mathrm{stC})\big(\boldsymbol{E}(\mathrm{stV/cm}) + \boldsymbol{v}(\mathrm{cm/s}) \times \boldsymbol{B}(\mathrm{statgauss})\big) \; . \tag{58}$$

The **resistance unit** is

$$1 \; \mathrm{st\Omega} \equiv 1\,\mathrm{stV}/1\,\mathrm{stA} = 1\,\mathrm{sec/cm},$$

with the dimensions of 1/velocity. Relative to SI, $1\,\mathrm{st\Omega} = 9\times10^{11}\,\Omega$. Dimensions of resistivity are sec, and conductivity is 1/sec. Relative to SI, 1 Siemens/m = $9\times10^9$ sec$^{-1}$.

The **capacitance unit** is

$$1 \; \mathrm{stFarad} \equiv 1\,\mathrm{stC}/1\,\mathrm{stV} = 1\,\mathrm{cm},$$

with the dimensions of length. Relative to SI, $1\,\mathrm{stF} = 1/9\times10^{11}\,\mathrm{F} = 1.11\times10^{-12}\,\mathrm{F}$.

Due partly to the inconvenient current and resistance units, like pure emu, pure esu is essentially never used today.

Even when discussing fundamental issues rather than laboratory-scale quantities, pure emu and pure esu are inconvenient, partly because of the $c^2$ factors in Maxwell's Equations, and those systems have given way to the Gaussian System.

## 6.3 THE <u>GAUSSIAN</u> SYSTEM

It was Heinrich Hertz who invented and named the ***Gaussian*** system (Lorentz [Lo04] credits Gauss, Helmholtz, and Hertz [Si62]). In the 1880's he was testing Maxwell's new theory and seeking electromagnetic waves when he realized a more useful unit system could be devised from the older esu and emu systems [Ne00].

This new method takes advantage of the finer aspects of esu and the finer aspects of emu. The units for electrical quantities are adopted from the esu system and units for magnetic quantities are taken from emu[62]. Like those two older systems, it is a cgs system, and is unrationalized.

For this scheme, then, it is clear that charge should be measured in esu (stC, Eq (50)). But the units and dimensions in which current should be measured are not so clear. Current could be considered an "electric" quantity (a moving charge) and so measured in esu (stA), or it could be considered a "magnetic" quantity (the source of $\boldsymbol{B}$) and so measured in emu (abA). To establish the Gaussian system one must choose current to be either an electric or magnetic quantity.

There is no "correct" choice. The more common choice is to consider current an electric quantity[63], so that the Gaussian unit of current is the stA. In the Gaussian system then, $k_e = 1$, $k_B = c$, and so $k_m = 1/c^2$, $k_F = 1/c$.

In Gaussian Units the force between two charges is taken from esu, (49). Since *I* is also in esu, the force between two currents is also the same as esu, (52), so that

$$f(\text{dyn}) = \frac{q_1 q_2}{r^2}$$

$$f'(\text{dyn/cm}) = 2\frac{1}{c^2}\frac{I_1 I_2}{r} \tag{59}$$

with $q$ in stC and $I$ in stA.

---

[62] In the literature of the early 20$^\mathrm{th}$ century, the Gaussian system is sometimes called the *symmetric* system.
[63] As stated by Birge [Bi34, p.44], and as used, e.g., in Jackson [Ja75] and in Landau and Lifschitz [La60].

Likewise, electric potential, field, and the continuity equation read the same as in esu (53), (54), and (55).

For $B$ of a wire current, Eq (10), $k_B k_m = 1/c$, so that

$$B = \frac{2}{c}\frac{I}{r} \quad \text{(Gaussian)} \tag{60}$$

with $I$ in stA. Since $I(\text{stA})/c = I(\text{abA})$, this is the same field as in emu, Eq (33). Thus the gauss can still be used as a unit of $B$; its definition in Gaussian Units parallels that in emu (37),

**1 gauss** $\equiv$ 1 Gaussian unit of magnetic induction = $(3 \times 10^{10}\text{cm/s})$ stA/cm = 1 dyn$^{1/2}$/cm. (61)

These dimensions of $B$ (stA/s) are the same as $E$ (stV/cm), namely dyn$^{1/2}$/cm, consistent with (11) with $k_B = c$. Note that in emu, no $c$ occurs in the field of a wire, Eq (33); in esu the factor is $1/c^2$, Eq (56); and in Gaussian Units, the factor $1/c$, Eq (60).

### 6.3.1 MAXWELL'S EQUATIONS IN GAUSSIAN UNITS

In the absence of material media, Maxwell's Equations in Gaussian Units are

$$\begin{aligned}
\nabla \cdot \boldsymbol{E} &= 4\pi\rho \\
\nabla \times \boldsymbol{B} - \frac{1}{c}\frac{\partial \boldsymbol{E}}{\partial t} &= \frac{4\pi}{c}\boldsymbol{j} \\
\nabla \times \boldsymbol{E} + \frac{1}{c}\frac{\partial \boldsymbol{B}}{\partial t} &= 0 \\
\nabla \cdot \boldsymbol{B} &= 0
\end{aligned} \tag{62}$$

The dimensions are:

$\boldsymbol{E}$ (stV/cm); $\quad\quad\quad \boldsymbol{B}$ (gauss = stV/cm)
$\rho$ (stC/cm$^3$); $\quad\quad \boldsymbol{j}$ (stA/cm$^2$); $\quad\quad \boldsymbol{j}/c$ (stC/cm$^3$)

In (62) factors of $c$ occur only to the first power; factors of $c^2$ in emu (48) and in esu (57) have disappeared. This is the simplification of the Gaussian system over emu and esu alluded to by Hertz.

With the constitutive relations,

$$\begin{aligned}
\boldsymbol{D} &= \boldsymbol{E} + 4\pi\boldsymbol{P} \\
\boldsymbol{H} &= \boldsymbol{B} - 4\pi\boldsymbol{M}
\end{aligned} \tag{63}$$

Maxwell's Equations in Gaussian Units in material media become

$$\nabla \cdot \boldsymbol{D} = 4\pi\rho$$
$$\nabla \times \boldsymbol{H} - \frac{1}{c}\frac{\partial \boldsymbol{D}}{\partial t} = \frac{4\pi}{c}\boldsymbol{j} \qquad (64)$$
$$\nabla \times \boldsymbol{E} + \frac{1}{c}\frac{\partial \boldsymbol{B}}{\partial t} = 0$$
$$\nabla \cdot \boldsymbol{B} = 0$$

where $\rho$ and $\boldsymbol{j}$ are free charge and current densities.

According to (12) with $k_B = c$, the **Lorentz force in Gaussian Units** reads

$$\boldsymbol{F}_L = q(\boldsymbol{E} + \frac{\boldsymbol{v}}{c} \times \boldsymbol{B}) , \qquad (65)$$

or

$$\boldsymbol{F}_L(\text{dyn}) = q(\text{stC})\left(\boldsymbol{E}(\text{stV/cm}) + \frac{\boldsymbol{v}(\text{cm/s})}{c(\text{cm/s})} \times \boldsymbol{B}(\text{gauss})\right) . \qquad (66)$$

The **force per unit length** $f'$ on a wire carrying current $I$ in a field $B$ is

$$f'(\text{dyn/cm}) = (I(\text{stA})/c)B(\text{gauss}) ,$$

the counterpart of emu's Eq (44).

The relation between the magnitudes $E$ and $B$ fields in a **plane wave in vacuum** is

$$E(\text{stV/cm}) = B(\text{gauss}) .$$

The **resistance unit** is

$$1 \text{ stOhm} \equiv 1 \text{ stV}/1 \text{ stA} = 1 \text{ sec/cm},$$

with dimensions of 1/velocity. Relative to SI, $1 \text{ st}\Omega = 9 \times 10^{11} \, \Omega$.

Resistivity has dimension st$\Omega$·cm = sec, and conductivity is 1/sec.

The **capacitance** has dimensions of length, and its **unit** is

$$1 \text{ stF} \equiv 1 \text{ stC}/1 \text{ stV} = 1 \text{ cm},$$

Relative to SI, $1 \text{ stF} = 1/(9 \times 10^{11}) \text{ F} \approx 1.11 \text{ pF}$.

The capacitance of parallel plates each of area $A$ (cm$^2$) separated by $d$ by a medium of relative dielectric constant $\epsilon_r$ is

$$C = \frac{\epsilon_r A}{4\pi d} \quad (\text{cm})$$

**Inductance** has dimensions of T$^2$/L; its unit is

$$1 \text{ stHenry} \equiv 1 \text{ stV·sec}/1 \text{ stA} = 1 \text{ sec}^2/\text{cm}$$

Relative to SI, $1 \text{ stHenry} = 9 \times 10^{11}$ Henry. The induced emf is

$$\text{emf}(\text{stV}) = L(\text{s}^2/\text{cm}) \cdot dI(\text{stA})/dt .$$

The **Poynting Vector** is

$$S(\text{erg/cm}^2/\text{sec}) = \frac{c}{4\pi} E(\text{stV/cm}) \times H(\text{oersted}) \qquad (67)$$

**Energy density** is

$$u = \frac{1}{8\pi}(\boldsymbol{B}\cdot\boldsymbol{H} + \boldsymbol{E}\cdot\boldsymbol{D}) \quad (\text{erg/cm}^3),$$

with **energy conservation**

$$\nabla\cdot\boldsymbol{S} + \frac{\partial u}{\partial t} = -\boldsymbol{j}\cdot\boldsymbol{E} \qquad (68)$$

as in emu, but with $j$ in stA/cm$^2$ and $E$ in stV/cm. In a **plane wave** in vacuum, $S = cu$. And in vacuum, since $E(\text{stV/cm}) = B(\text{gauss}) = H(\text{oersted})$, the **impedance of free space** is $E/H = 1$.

**The Fine Structure Constant**

Since $\alpha$ involves only electrostatic factors, its appearance when written in terms of $e$, $\hbar$, and $c$ is the same in esu or Gaussian units (but different from emu). It appears as

$$\alpha = e^2/\hbar c = e(\text{stC})^2/\hbar(\text{erg·s})c(\text{cm/s}) = 1/137 \qquad (\text{esu, Gaussian}).$$

## 6.4 THE <u>VARIANT-GAUSSIAN</u> SYSTEM

In the Gaussian system, current was chosen to be an electric quantity, and so measured in stA. A *variant of the Gaussian* system results by choosing current to be a magnetic quantity. Then in the variant system the basic unit of current is the abA. Such a choice may be desirable because the abA ($= 10\,\text{A} = 3\times 10^{10}\,\text{stA}$) may be a more convenient current unit than the much smaller stA.

In the variant system, charge is the same as in esu (stC), and the electrostatic force (3) is likewise taken as in esu, (49), so $k_e = 1$. The magnetostatic force (4) is taken as in emu, (27), so $k_m = 1$. While in Gaussian units the two forces are as in (59), in Variant-Gaussian they are

$$f(\text{dyn}) = \frac{q_1 q_2}{r^2}$$

$$f'(\text{dyn/cm}) = 2\frac{I_1 I_2}{r} \qquad (69)$$

with $q$ in stC and $I$ in abA.

Then the ratio of forces $f/f'$ becomes, instead of Eq (6),

$$\frac{f}{f'} \sim \text{L} \sim \frac{q_1 q_2 / \text{L}^2}{I_1 I_2 / \text{L}} \sim \frac{q^2}{I^2}\frac{1}{\text{L}}, \qquad (70)$$

which means

$$[q] = [I \cdot L] = c[I \cdot t] \qquad \text{(Variant Gaussian)} \tag{71}$$

instead of the more common (5). Choosing $k_e = 1$ and $k_m = 1$ violates constraint (7) and forces charge and current to be related by $q/I \sim$ length instead of time. Constraint (7) is replaced by $k_e/k_m = 1$. Thus Maxwell's Equations in Variant-Gaussian units do not conform to form (20) with any choice of $k_B$. V-G is a nonconforming rogue convention.

In view of (71), the continuity equation reads

$$\nabla \cdot \boldsymbol{j} + \frac{1}{c}\frac{\partial \rho}{\partial t} = 0 \qquad \text{(Variant Gaussian)} \tag{72}$$

instead of the usual (8). Thus $\boldsymbol{j}$ and $\rho$ have the same dimensions, for $\boldsymbol{j}$ (abA/cm$^2$) = $(1/c)$ $\boldsymbol{j}$(stA/cm$^2$) = $(1/c)\boldsymbol{j}$(stC/s·cm$^2$) has the dimension stC/cm$^3$.

In Variant-Gaussian Units, charge, electric field, and potential are in stC, stV/cm, and stV respectively, as in esu.

The magnetic field is in gauss and is defined as in emu, Eq (37). The magnetic induction of a wire is

$$\boldsymbol{B}(\text{gauss}) = 2\frac{I(\text{abA})}{r} \qquad \text{(Variant-Gaussian, and emu)} \tag{73}$$

the same as (33). Maxwell's Equations in the "most general" form (16) assumed no constraints among $k_e$, $k_m$, $k_B$, and $k_F$, but they did assume the usual form (8) of the continuity equation. As that has been replaced by (72) there are no values for $k_e$, $k_m$, $k_B$, and $k_F$ for which Eqs (16) are the correct Maxwell's Equations in Variant-Gaussian units. Rather, we have:

### 6.4.1 MAXWELL'S EQUATIONS IN VARIANT-GAUSSIAN UNITS

In the absence of material media, Maxwell's Equations are

$$\begin{aligned}
\nabla \cdot \boldsymbol{E} &= 4\pi\rho \\
\nabla \times \boldsymbol{B} - \frac{1}{c}\frac{\partial \boldsymbol{E}}{\partial t} &= 4\pi \boldsymbol{j} \\
\nabla \times \boldsymbol{E} + \frac{1}{c}\frac{\partial \boldsymbol{B}}{\partial t} &= 0 \\
\nabla \cdot \boldsymbol{B} &= 0
\end{aligned} \tag{74}$$

The dimensions of variables are $\boldsymbol{E}$(stV/cm), $\boldsymbol{B}$(gauss), $\rho$(stC/cm$^3$), $\boldsymbol{j}$(abA/cm$^2$). Relative to Gaussian units, Eqs (62), only the rhs of Ampere's Law has changed, from $(4\pi/c)\boldsymbol{j}$ to $4\pi\boldsymbol{j}$, with $\boldsymbol{j}$ now in abA/cm$^2$. Eqs (74) are appealing in that $c$ occurs only multiplying $\partial t$, and $c$ disappears altogether if one uses the time variable $\tau = ct$.

Purists may say variant-Gaussian is a mix of unit systems, and does not qualify as a separate system of units[64].

---

[64] Jackson [Ja75, Appendix] refers to Variant-Gaussian as modified Gaussian, and does not accord it the status of a stand-alone unit system. Panofsky and Phillips [Pa62], however, refer to our variant Gaussian

The variant Gaussian system is occasionally used today when the magnitude of its current unit is convenient (e.g., by some work in plasma physics [Lo63] and in nuclear EMP [Lo78]), but not as frequently as the standard Gaussian system.

Confusingly, sometimes the variant-Gaussian system is referred to as "Gaussian units" [e.g., Lo63, Lo78, Pa62].

The **Lorentz force** in Variant-Gaussian Units reads the same as in Gaussian units,

$$F_L(\text{dyn}) = q(\text{stC})\left(E(\text{stV/cm}) + \frac{v(\text{cm/s})}{c(\text{cm/s})} \times B(\text{gauss})\right), \tag{75}$$

since $q$, $E$, and $B$ are in the same units as in Gaussian. The stV has the same dimensions as abA, namely $\text{dyn}^{1/2}$.

The **force per unit length** $f'$ on a wire carrying current $I$ in a field $B$ is

$$f'(\text{dyn/cm}) = I(\text{abA})B(\text{gauss}),$$

the same as emu Eq (44).

The relation between the magnitudes $E$ and $B$ fields in a **plane wave in vacuum** is

$$E(\text{stV/cm}) = B(\text{gauss}),$$

the same as Gaussian.

**Resistance**, in Variant-Gaussian Ohms, is

$$R(\text{VG}\Omega) \equiv V(\text{stV})/I(\text{abA}) = cV(\text{stV})/I(\text{stA}) = \text{dimensionless}.$$

The *unit* of resistance is equal to $1\,\text{stV}/1\,\text{abA}$. Relative to SI, the V-G unit is the same as $(300\text{V})/(10\text{A}) = 30\,\Omega$. Resistivity $\rho$ has dimension cm, and conductivity $\sigma$ is $\text{cm}^{-1}$. Relative to other systems,

$$1\,\text{S/m (SI)} = 9\times10^9\,\text{sec}^{-1}\,(\text{esu, Gaussian}) = 0.3\,\text{cm}^{-1}\,(\text{VG}).$$

The **capacitance unit** is the same as in esu:

$$1\,\text{stFarad} \equiv 1\,\text{stC}/1\,\text{stV} = 1\,\text{cm},$$

with dimensions of length.

The **capacitance** of parallel plates of area $A$ separated by $d$ with a material of relative permeability $\epsilon_r$ is the same as esu and Gaussian:

$$C = \frac{\epsilon_r A}{4\pi d} \quad (\text{cm})$$

and has the dimension of length. It is $A/4\pi d$ in vacuum.

**Inductance $L$** can be defined either by

---

system as the Gaussian system [Appendix I, p. 461, and Table I-2]. As their text uses SI, the naming convention is not elaborated further. Longmire [Lo63] also calls this system Gaussian units.

$$V = L\frac{dI}{dt} \tag{76}$$

in which case $L$ has dimensions stV·sec/abA = time, or by

$$V = L\frac{dI}{cdt} \tag{77}$$

in which case $L$ has dimension length. There does not appear to be a universally accepted preference.

The **Poynting Vector** is

$$S(\text{erg/cm}^2/\text{sec}) = \frac{c}{4\pi} E(\text{stV/cm}) \times H(\text{oersted}) \tag{78}$$

**Energy density** is

$$u = \frac{1}{8\pi}(\boldsymbol{B} \cdot \boldsymbol{H} + \boldsymbol{E} \cdot \boldsymbol{D}) \quad (\text{erg/cm}^3),$$

with **energy conservation**

$$\nabla \cdot \boldsymbol{S} + \frac{\partial u}{\partial t} = -\boldsymbol{j} \cdot \boldsymbol{E} \tag{79}$$

the same as Gaussian units. In a **plane wave** $E = B$, and $S = cu$.

Since the variant Gaussian system also uses the esu unit of charge, stC, the **fine structure constant** also appears written as $\alpha = e^2/\hbar c = 1/137$.

### 6.5 "RATIONALIZING" A UNIT SYSTEM

In all systems discussed so far, factors of $4\pi$ occur in Maxwell's Equations in front of $\rho$ and $\boldsymbol{j}$, and do not occur in the force laws adapted from (3) and (4). To Heaviside, this was quite unnatural, even irrational. He fought to choose units in which the denominator was $4\pi r^2$ in the force law for problems of spherical symmetry, or $2\pi r$ in cylindrical symmetry.

Heaviside wanted units such that the Coulomb force law would read

$$f = \frac{q_1 q_2}{4\pi r^2} \tag{80}$$

or at least

$$f = \text{constant} \times \frac{q_1 q_2}{4\pi r^2} \tag{81}$$

instead of Eq (49) without the $4\pi$. This form, (charge)$^2$/(area of a sphere), displays that the electric flux is conserved through the sphere area $4\pi r^2$. Then Gauss' Law will read

$$\nabla \cdot \boldsymbol{E} = \rho \tag{82}$$

instead of with $4\pi$ multiplying $\rho$ as in Eq (57). The denominator of the force law is where the factor $4\pi$ "belongs", he said. A system with that convention is now called *Rationalized* (a possible play on words). Thus esu, emu, and Gaussian are unrationalized systems; Heaviside-Lorentz and SI, to be discussed, are rationalized.

A factor $4\pi$ enters at all because of the way Coulomb's law relates the electric field $E$ (force per unit charge) outside a volume to the charge $Q$ in the volume. It is fundamentally because the surface area of a sphere is $4\pi r^2$.

When you rationalize Gaussian Units, the result is called Heaviside-Lorentz Units. To obtain SI, two major changes are made. First, a fourth base dimension, current, is added, along with its unit, the Ampere. And second, constants $k_e$ and $k_B$ are chosen to take $4\pi$'s out of Maxwell's Equations and put them in the denominator of the force laws.

## 6.6  THE HEAVISIDE-LORENTZ SYSTEM

As mentioned, the Gaussian system is preferred over SI for fundamental discussions. But being unrationalized, there are inconvenient $4\pi$'s in Maxwell's Equations. To correct this, another system has been devised to rationalize the Gaussian system and eliminate those factors. H. A. Lorentz (1853-1928) was a prominent physicist at the time (and one of Einstein's idols), and favored Gaussian units *and* Heaviside's rationalization ideas. He combined the two as follows. Instead of the first line of (59), the point force law is written as in (80) with electric field $E = q/4\pi r^2$. Then Gauss' Law reads as Eq (82).

The resulting system is called the **_Heaviside-Lorentz_** system (also called the **_Rationalized Gaussian_** system), which is used today mostly in quantum field theory. The advantage is that field theorists are not bothered by factors of $4\pi$ in the field Lagrangian.

The electron charge expressed in esu is $e(\text{esu}) = 4.803\times 10^{-10}$ stC. What is it when expressed in HL units? The force between two electrons separated by $r$ is

$$f = \frac{e(\text{esu})^2}{r^2} = \frac{e(\text{HL})^2}{4\pi r^2} \tag{83}$$

so that the electron charge in HL units is $e(\text{HL}) = \sqrt{4\pi} \times 4.803\times 10^{-10} = 1.703\times 10^{-9}$. Thus, while 1 unit of charge in esu is 1 stC, 1 unit of charge in HL units is $1/\sqrt{4\pi}$ stC $\approx 0.282$ stC.

We are unaware of any agreed upon symbol for the Heaviside Lorentz unit of charge, but we will use **hC** for it. Thus

$$1\,\text{hC} = \text{Unit of charge in HL units} = \frac{1}{\sqrt{4\pi}}\,\text{stC} \approx 0.2821\,\text{stC} \tag{84}$$

The fine structure constant is written as $\alpha = e^2/4\pi\hbar c = 1/137$, with $e$ in hC.

In Heaviside-Lorentz Units, continuity reads

$$\nabla\cdot\boldsymbol{j} + \partial\rho/\partial t = 0,\ \text{with}\ \boldsymbol{j}\ \text{in hC/sec}\cdot\text{cm}^2,\ \text{and}\ \rho\ \text{in hC/cm}^3.$$

### 6.6.1 MAXWELL'S EQUATIONS IN HEAVISIDE-LORENTZ UNITS

In the absence of material media, Maxwell's Equations are

$$\nabla \cdot \boldsymbol{E} = \rho$$
$$\nabla \times \boldsymbol{B} - \frac{1}{c}\frac{\partial \boldsymbol{E}}{\partial t} = \frac{1}{c}\boldsymbol{j} \quad (85)$$
$$\nabla \times \boldsymbol{E} + \frac{1}{c}\frac{\partial \boldsymbol{B}}{\partial t} = 0$$
$$\nabla \cdot \boldsymbol{B} = 0$$

$\rho$ is in units of (hC/cm$^3$), $\boldsymbol{j}$ in (hC/s·cm$^2$), $\boldsymbol{E}$ in ($E_h$), and $\boldsymbol{B}$ in gauss, the same as the Gaussian (62) with all $4\pi$'s replaced by 1.

### 6.7 THE <u>SI</u> SYSTEM

That esu and emu resistance, current, and potential units were inconvenient was a prime motivating factor driving the creation of another more practical system of units, first suggested by Giorgi in the early 1900's, that eventually grew into present day *SI*.

An official unified international system of units (for electromagnetism, thermodynamics, and all of physics) was finally adopted in 1971 (and included the mole) and recommended for universal use. This is the *Système International*, or <u>**SI**</u> system of units. It adopts the mechanical units kilogram, meter, and second. The basic physical force used to connect electromagnetic phenomena to mechanical phenomena is taken to be the magnetostatic force between two long, parallel, current carrying wires[47], as in emu. The Coulomb force between two point charges is <u>not</u> used to define this system.

Experimentally the force per unit length between wires separated by *r* is proportional to 1/*r* and to the product $I_1 I_2$ of their currents. The current is the "strength" of the source of the force, and the force law is written as in Eq (4),

$$f'(\text{N/m}) = 2k_m \frac{I_1 I_2}{r} \quad (86)$$

Unlike all previous unit systems discussed above, in SI a new base dimension in addition to mass, length, and time is introduced that is of a distinctly electromagnetic, non-mechanical, nature. (A comparable new electromagnetic dimension had been introduced in Giorgi's MKSQ system; SI is an outgrowth of that system). As a result, the parameter $k_m$ in (86) itself has dimensions that includes both mechanical and current dimensions.

The new dimension is current, and its unit is called the Ampere. One needs to define its magnitude, and then specify $k_m$. Recall that in emu the unit of current is that which causes a force of 2 dyn/cm between two wires separated by 1 cm, Eq (27). To maintain the main features of the long-used emu and Gaussian systems, 1 Ampere (1 A) was defined to be both a value convenient for laboratory work and simply related to the abA. The value

**1 A = 0.1 abA**

satisfies both conditions. Then the force between two wires separated by $r$ each carrying current $I$ is, from (27),

$$f'(\frac{\text{dyn}}{\text{cm}}) = 2\frac{I(\text{abA})^2}{r(\text{cm})}$$
$$= 10^3 f'(\frac{\text{N}}{\text{m}}) = 2\frac{(0.1I(\text{A}))^2}{100\, r(\text{m})} = 2\times 10^{-4} \frac{I(\text{A})^2}{r(\text{m})}, \qquad \text{or} \qquad (87)$$
$$f'(\frac{\text{N}}{\text{m}}) = 2\times 10^{-7} \frac{I(\text{A})^2}{r(\text{m})}$$

so that the force between two wires separated by 1 m, each carrying 1 A, is $2\times 10^{-7}$ N/m. That can be taken as the definition of 1 Ampere.[65] In deference to Rationalization, to display a $4\pi$ in the denominator, the proportionality factor $k_m$ is traditionally written in terms of a different constant $\mu_o$ as $\boldsymbol{k_m = \mu_o/4\pi}$ so that[66, 67]

$$f' = 2k_m \frac{I^2}{r} = 2\frac{\mu_o}{4\pi}\frac{I^2}{r} = \frac{\mu_o}{2\pi}\frac{I^2}{r} \tag{88}$$

and (87) immediately requires $\mu_o$ have magnitude $4\pi\times 10^{-7}$. With distance in meters, current in Amperes, and force per unit length in Newton/meter, the proportionality constant $\mu_o$ is in N/A$^2$. (The N·m/A$^2$ = V·s/A = $\Omega$·s is called the Henry, H, a unit of inductance. Therefore $\mu_o$ is in H/m). Thus the proportionality constant $\mu_o$ must have the exact value $4\pi\times 10^{-7}$ H/m. Its value is the result of the chosen definition of the unit for current. With the $4\pi$ in the denominator of the force law (or $2\pi$ for line forces), SI is a *rationalized* system. (One may accuse SI of legerdemain; while $4\pi$ has been put in the denominator, it is just taken out again in the value of the new constant $\mu_o$. Heaviside wanted a $4\pi$ to explicitly *appear* in the denominator.)

Prior to the introduction of SI, the most commonly used system was electromagnetic units. When SI was established, for simplicity units for most quantities (e.g., resistance, current, potential, …) were chosen to be both convenient on the laboratory scale, and to be simply

---

[65] The international governing body, the Bureau International des Poids et Mesures (BIPM), is considering updating the definitions of various SI units. See http://www.bipm.org/en/si/new_si/ and [Ne14].

[66] The factor 2 arises as follows. The force between two infinitesimal current elements $i_1 ds_1$ and $i_2 ds_2$ is $df = (\mu_o/4\pi)\boldsymbol{i}_1\times(\boldsymbol{i}_2\times\boldsymbol{r}_{12})ds_1 ds_2/r_{12}^3$, and may be considered more fundamental than the force between two long wires. When one integrates this expression to obtain the force per unit length between long wires, the integral results in a factor of 2. It turns the $1/4\pi r^2$ point-point law into the $1/2\pi r$ line-line law. This reason applies, of course, to all units systems.

[67] A different factor 2, separate from that in fn 66, has an historical twist. In Ampere's day there was an active theory, associated with the name Fechner (previously suggested by Coulomb and Abbe Nollet in the 18$^{th}$ century), that current in conductors is composed of two electrical fluids, of opposite charge and flowing in opposite directions. This left an ambiguity of a factor of two according to the meaning of "current", and led to two different conventions [As04]. It has been suggested that this factor may have confused Weber and Kohlrausch in interpreting their measurements of the ratio of emu to esu charge units [As03].

Curiously, the unit of current in Ampere's *Electrodynamic Units* was $\sqrt{2}$ larger than in emu. In those units the factor 2 in Eq (27) would be 1. The *Electrodynamic Units* system was abandoned.

related to their emu counterparts, by an integer power of ten. The SI unit of a quantity was chosen to be that power of 10 times its emu unit that resulted in a laboratory-convenient size [Ko86, St41 (Section 1.8)]. Hence the definition of the Ampere. That makes the force $f'$ come out as stated. The sole reason for the $10^{-7}$ in the definition is so that $f'$ will be the same as emu at the same current and the same distance. These definitions were chosen to minimize trouble to physicists transitioning from the familiar, widely used emu to the new SI, and still provide units that were neither too large nor too small. Existing equipment would not have to be recalibrated. This would be another reason SI was based on the magnetic force and not the point charge electrostatic force, since emu was based that way. Actually, the Ampere was first defined as 0.1 abA in the first (1881) IEC meeting in Paris, and the definition remained suitable for the much later SI.

As electric charge is current×time, the unit of charge is fixed by those of current and time; it is called the *Coulomb* (C), and $1\,C \equiv 1\,A \times 1\,sec$. Relative to previous units, $1\,C = 3 \times 10^9$ stC = 0.1 abC. The electron charge is $1.602 \times 10^{-19}$ C.[68] Appealing to (7), $k_e = c^2 k_m = c^2 \mu_o/4\pi$, and the force between two point charges $q_1$ and $q_2$ in vacuum, each in Coulombs, comes out

$$f = \frac{\mu_o}{4\pi} c^2 \frac{q_1 q_2}{r^2} \equiv \frac{q_1 q_2}{4\pi \epsilon_o r^2} \tag{89}$$

in Newtons[69]. The force between two electrons separated by 1 cm is $(4\pi \times 10^{-7}/4\pi)(3 \times 10^8)^2 \times (1.602 \times 10^{-19})^2 / (0.01\,m)^2 = 2.31 \times 10^{-24}$ N $= 2.31 \times 10^{-19}$ dyn, as in esu or emu of course[70].

The relation $1/\epsilon_o \mu_o =$ (velocity)$^2$ of electromagnetic waves follows directly from Maxwell's equations, so $\epsilon_o$ in the vacuum force law is **defined by** $\epsilon_o \mu_o = 1/c^2$ ($c = 3 \times 10^8$ m/s). Thus $\mu_o$ is set by definition, and $\epsilon_o = 1/c^2 \mu_o$ is set by $c$. In SI the velocity $c$ itself does not appear directly in any of Maxwell's Equations, only $\epsilon_o$ and $\mu_o$ appear.

From (89) the potential energy between two charges is

$$\Phi = \frac{q_1 q_2}{4\pi \epsilon_o r} \quad (J) \tag{90}$$

so that the potential of a single charge is

$$V = \frac{q}{4\pi\epsilon_o r} \quad (\text{J/C} = \text{V}) \tag{91}$$

---

[68] Since the Coulomb and Ampere are so simply related in SI units, this system has in the past been called the MKSQ system (or Giorgi MKS system). Stratton (St41) and Sommerfeld (So52) call it that and use it (Stratton, §1.8, and Appendix I; Sommerfeld, Part I).

[69] If one replaces $4\pi\epsilon_o$ in (89) by 1, one regains the electrostatic force law in esu (49). This replacement displays the practicing physicist's dangerous rule of thumb that to change an equation from SI to esu, just replace $4\pi\epsilon_o$ by 1. The rule must be used with much caution; it does not always work, for example in the constitutive relations among $D$, $E$, and $P$.

[70] George Johnstone Stoney is credited as the first to suggest that electricity in matter is carried by atom-size units. In 1891 he suggested the unit be called the *electron*. Linus Pauling has suggested that in Stoney's honor an appropriate unit of charge in the SI system, rather than the Coulomb, would be that amount such that two of them 1 meter apart exert a repulsive force of 1 Newton; he suggested it be called the *stoney* [Pa70b]. 1 stoney = $1.0548 \times 10^{-5}$ Coulomb. Evidently the suggestion has not gained wide acceptance.

The electric field of a point charge is

$$E = -\frac{\partial V}{\partial r} = \frac{q}{4\pi\epsilon_o r^2} \quad (\text{V/m}) \tag{92}$$

**Magnetic Induction $B$**

From (10) and with $k_m = \mu_o/4\pi$, the magnetic induction $B$ is defined as

$$B = k_B \frac{\mu_o}{2\pi}\frac{I}{r} = \frac{\mu_o}{2\pi}\frac{I}{r} = 2\times 10^{-7}\frac{I}{r} \tag{93}$$

with $I$ in A, $r$ in m, and $\mu_o$ in H/m. The equation indicates that the choice made in SI is $k_B = 1$ (Table 1).

The **unit** of magnetic induction is the **tesla** (T) alternately defined (from (93)) as

- the field 1 m from a wire current of $5\times 10^6$ A, or as
- the magnetic induction that generates a potential of one volt in a conductor of length one meter when moving at a rate of one meter per second.

Its units are $[B] = \text{H·A/m}^2 = \text{V·s/m}^2 = [E/c]$. Compare with gauss, Eq (33) *ff*. In SI the gauss does not occur. 1 tesla is a large field. 1 T = $10^4$ gauss ≈ 20,000×geomagnetic field ≈ the field used in MRI scanners. The superconducting magnets in CERN's Large Hadron Collider produce about 8 T, requiring magnet wire currents of order 12 kA with many turns.

**Magnetic flux** is in tesla·m². 1 T·m² ≡ 1 **weber** (W). So $B$ can be stated in W/m².

In SI the **magnetic field strength $H$** is the field $B/\mu_o$ (in vacuum), with units A/m. In SI the oersted does not occur.

### 6.7.1 CONSTITUTIVE RELATIONS

From Eq (21), in SI one has $D$ and $H$ defined as

$$\begin{aligned} D &= \epsilon_o E + P \\ H &= \frac{1}{\mu_o}B - M \end{aligned} \tag{94}$$

with dimensions $[D] = [P] = \text{C/m}^2$, $[E] = \text{V/m}$; $[H] = [M] = \text{A/m}$; $[B] = \text{T}$[71].

### 6.7.2 MAXWELL'S EQUATIONS IN SI

Maxwell's Equations in SI in material media become

---

[71] As of 1973 there was still debate about the definition of $H$ in terms of $B$ [Co73]. Folks settled on Eq (94).

$$\nabla \cdot \boldsymbol{D} = \rho$$
$$\nabla \times \boldsymbol{H} - \frac{\partial \boldsymbol{D}}{\partial t} = \boldsymbol{j} \tag{95}$$
$$\nabla \times \boldsymbol{E} + \frac{\partial \boldsymbol{B}}{\partial t} = 0$$
$$\nabla \cdot \boldsymbol{B} = 0$$

where $\rho$ and $\boldsymbol{j}$ are free charge and current densities. An appealing aspect of SI is that, of all the units systems, Maxwell's Equations appear in their simplest form; no numerical coefficients occur multiplying terms.

### 6.7.3 OTHER DYNAMICAL QUANTITIES

According to (12), in SI the **Lorentz force** reads

$$\boldsymbol{F}_L = q(\boldsymbol{E} + \boldsymbol{v} \times \boldsymbol{B}) \ . \tag{96}$$

or, with units spelled out:

$$\boldsymbol{F}_L(N) = q(C)\bigl(\boldsymbol{E}(V/m) + \boldsymbol{v}(m/s) \times \boldsymbol{B}(T)\bigr) , \tag{97}$$

confirming that the **force per unit length** $f'$ on a wire carrying current $I$ in an external field $B$

$$f'(N/m) = IB \ , \tag{98}$$

with $I$ in A and $B$ in tesla.

In SI the **continuity** law reads the conventional way,

$$\nabla \cdot \boldsymbol{j} + \frac{\partial \rho}{\partial t} = 0 \tag{99}$$

now with $\boldsymbol{j}$ in A/m$^2$, and $\rho$ in C/m$^3$.

In a **plane wave in vacuum**

$$E(V/m) = c(m/s)B(T) = 3 \times 10^8 B(T) \ .$$

An electric field $E$ of $3 \times 10^4$ V/m accompanies a magnetic induction $B$ of $10^{-4}$ T (= 1 gauss) or a magnetic field $H = B/\mu_o = 1000/4\pi$ A/m = 79.6 A/m.

The **resistance unit** is

$$1 \text{ Ohm} \equiv 1 \text{ V}/1\text{A},$$

**Resistivity** is in $\Omega \cdot$m, conductivity is in Siemens (S)/m (formerly mho/m).

The **Impedance of Free Space** $Z_o$ is

$$Z_o = E(V/m)/H(A/m) = \mu_o E(V/m)/B(T) = \mu_o c = \sqrt{\frac{\mu_o}{\epsilon_o}} = 377 \ \Omega$$

The **capacitance unit** is 1 Farad ≡ 1 F ≡ 1C/1V, and the capacitance of parallel plates of area $A$ (m$^2$), separation $d$ (m) separated by material of relative dielectric constant $\epsilon_r$ is

$$C = \frac{\epsilon_r A}{d} \quad (F)$$

The **Poynting Vector** is

$$\mathbf{S} = \mathbf{E} \times \mathbf{H} = \mathbf{E}(V/m) \times \mathbf{H}(A/m) \quad (W/m^2)$$

**Energy density** is

$$u = \frac{1}{2}(\mathbf{B} \cdot \mathbf{H} + \mathbf{E} \cdot \mathbf{D}) \quad (J/m^3),$$

with **energy conservation**

$$\nabla \cdot \mathbf{S} + \frac{\partial u}{\partial t} = -\mathbf{j} \cdot \mathbf{E} \tag{100}$$

with $j$ in A/m$^2$ and $E$ in V/m.

In SI the dimensionless **fine structure constant** appears as $\alpha = e^2/4\pi\epsilon_o \hbar c = 1/137$.

In previous unit systems, fractional powers of M, L, and T occur, for example Eqs (28) or (50). But when current (or charge, as in Giorgi MKSQ) is introduced as a fourth unit, the dimensions of dynamical quantities involve only integer powers of M,L,T, and A (or M,L,T, and Q). This may be considered another advantage of the SI system.

It seems that most (but not all) undergraduate level physics textbooks and engineering texts employ SI units, while most graduate physics texts use Gaussian units.

### 6.8 PHYSICAL PROPERTIES OF THE ELECTRON

After this foray through half a dozen systems, our feet can be put back on familiar ground by noting physical properties of the electron in the various systems:

The *electron charge* is

| | | | |
|---|---|---|---|
| $e =$ 1.602×10$^{-19}$ | Coulomb | | [SI units] |
| 4.803×10$^{-10}$ | stC | | [esu, Gaussian, and variant-Gaussian] |
| 1.602×10$^{-20}$ | abC | | [emu] |
| 1.703×10$^{-9}$ | hC ($=\sqrt{4\pi} \times 4.803\times 10^{-10}$) | | [H-L units] |

The *force f* between two electrons separated by $r = 1$ cm is $2.31 \times 10^{-19}$ dyn, and appears as

$$\begin{aligned}
f &= e^2/4\pi\epsilon_o r^2 &&= (1.602\text{E}{-19})^2/4\pi(8.854\text{E}{-12})(0.01\text{m})^2 \times 1\text{E}5(\text{dyn/N}) &&\text{[SI]} \\
&\phantom{=} e^2/r^2 &&= (4.803\text{E}{-10})^2/(1\text{cm})^2 &&\text{[esu, Gaussian, V-G]} \\
&\phantom{=} c^2 e^2/r^2 &&= (3\text{E}10)^2(1.602\text{E}{-20})^2/(1\text{cm})^2 &&\text{[emu]} \\
&\phantom{=} e^2/4\pi r^2 &&= (1.703\text{E}{-9})^2/4\pi(1\text{cm})^2 &&\text{[H-L]}
\end{aligned}$$

The *fine structure constant* α is always 1/137, and appears as

$$\begin{aligned}
\alpha &= e^2/4\pi\epsilon_o \hbar c &&\text{[SI]} \\
&\phantom{=}\ e^2/\hbar c &&\text{[esu, Gaussian, v-G]} \\
&\phantom{=}\ c^2 e^2/\hbar c &&\text{[emu]} \\
&\phantom{=}\ e^2/4\pi\hbar c &&\text{[H-L]}
\end{aligned}$$

The *classical electron radius* $r_e$ is $2.818\times10^{-13}$ cm, and appears as

$$\begin{aligned}
r_e &= e^2/4\pi\epsilon_o mc^2 = (1.602\text{E}{-}19)^2/4\pi(8.854\text{E}{-}12)(0.911\text{E}{-}30\text{kg})(3\text{E}8)^2 &&\text{[SI]} \\
&\phantom{=}\ e^2/mc^2 \ \ = (4.803\text{E}{-}10)^2/(0.911\text{E}{-}27\text{g})(3\text{E}10)^2 &&\text{[esu, Gaussian, var-G]} \\
&\phantom{=}\ (c^2 e^2)/mc^2 = e^2/m = (1.602\text{E}{-}20)^2/(0.911\text{E}{-}27\text{g}) &&\text{[emu]} \\
&\phantom{=}\ e^2/4\pi mc^2 = (1.703\text{E}{-}9)^2/4\pi(0.911\text{E}{-}27\text{g})(3\text{E}10)^2 &&\text{[H-L]}
\end{aligned}$$

# 7 A MAGNETIC EXAMPLE

To illustrate how the different systems give the same result, but different intermediate factors, consider the simple problem of a magnetic field terminating on a conducting surface, analyzed in three unit systems

Magnetostatics is controlled by Ampere's Law:

| | | | |
|---|---|---|---|
| Gaussian: | $\nabla \times \boldsymbol{H} = (4\pi/c)\boldsymbol{J}$ | [$\boldsymbol{H}$ (Oe), $\boldsymbol{J}$ (stA/cm$^2$)] |
| emu or variant Gaussian: | $\nabla \times \boldsymbol{H} = 4\pi\boldsymbol{J}$ | [$\boldsymbol{H}$ (Oe), $\boldsymbol{J}$ (abA/cm$^2$)] |
| SI: | $\nabla \times \boldsymbol{H} = \boldsymbol{J}$ | [$\boldsymbol{H}$ (A/m), $\boldsymbol{J}$ (A/m$^2$)] |

The magnetic field against a conducting surface vanishes *in* the conductor, and is non-zero outside. Boundary conditions on the tangential component $H_t$ of $\boldsymbol{H}$ are

| | | |
|---|---|---|
| Gaussian: | $H_t = (4\pi/c)K$ | [$H_t$ (Oe); $K$ (stA/cm)] |
| emu or variant Gaussian: | $H_t = 4\pi K$ | [$H_t$ (Oe); $K$ (abA/cm)] |
| SI: | $H_t = K$ | [$H_t$ (A/m); $K$ (A/m)] |

The discontinuity is, of course, maintained by the surface current $K$ (current per unit length).

As an example, select $K = 10$ A/m $= 0.01$ abA/cm $= 3\times10^8$ stA/cm. The magnetic field $H$ is 10 A/m in all cases, expressed in different units as

| | Skin Current | Magnetic Field |
|---|---|---|
| Gaussian: | $3\times10^8$ stA/cm | $4\pi\times0.01$ Oe |
| emu or variant Gaussian: | 0.01 abA/cm | $4\pi\times0.01$ Oe |
| SI: | 10 A/m | 10 A/m |

All physical values for $K$ and $H$ are, of course, equal.

# 8  UNITS USED IN SOME TEXTBOOKS

- Abraham & Becker, *The Classical Theory of Electricity and Magnetism* Blackie (1937)     Gaussian
- Stratton, *Electromagnetic Theory*, McGraw Hill (1941)     MKSQ (essentially SI)
- Sommerfeld, *Electrodynamics*, Academic Press (1952)     MKSQ (essentially SI)
- Landau and Lifshitz, *Electrodynamics of Continuous Media*, Pergamon (1960)     Gaussian
- Panofsky and Phillips, *Classical Electricity and Magnetism*, Addison-Wesley, 2nd ed. (1962)     SI
- C. L. Longmire, *Elementary Plasma Physics*, Wiley Interscience (1963)     Variant Gaussian
- Lorrain and Corson, *Electromagnetic Fields and Waves*, 2nd ed., W.H. Freeman (1970)     SI
- R. Becker, *Electromagnetic Fields and Interactions* (Blaisdell, 1964; Dover, 1982)     Gaussian
- J. van Bladel, *Electromagnetic Fields*, revised printing, Hemisphere Publ. (1985)     SI
- Krauss, *Antennas*, 2nd edition, McGraw Hill (1988)     SI
- W. R. Smythe, *Static and Dynamic Electricity*, 3rd ed., Hemisphere Publ. (1989)     SI
- Born & Wolf, *Principles of Optics*, 7th edition, Cambridge U. Press (1999)     Gaussian
- J. D. Jackson, *Classical Electrodynamics*, J. Wiley.
    - 1st edition (1962)     Gaussian
    - 2nd ed (1975)     Gaussian
    - 3rd ed (1999):
        - Chs.1–10     SI
        - Chs.11–16     Gaussian
- D. J. Griffiths, *Introduction to Electrodynamics*, 4th ed., Pearson Education, Inc. (2013)     SI

# 9 APPENDIX A

## GAUSS' ABSOLUTE MEASUREMENT OF THE GEOMAGNETIC FIELD

### INTRODUCTION

Before 1832 there were standard units for mechanical quantities, but not for electric or magnetic quantities. No one had a quantitative measure for charge or current or any electromagnetic quantity. The Ampere or Farad, etc., did not exist.

At that time the geomagnetic field was a subject of interest, especially to the maritime community. The time-honored method of measurement was to use a compass needle constrained to rotate in a horizontal plane; it would line up with (the horizontal component of) **B** and measure its direction east or west of true (geographic) north (this angle is the *declination* of the geomagnetic field). If held to rotate in a vertical plane, it would measure the field's local dip angle (*inclination*). These angles were, of course, absolute measures. But the only measures of the magnitude of **B** were relative. One typically measured the oscillation period of a needle; the faster it oscillated the larger **B**. And one had to hope that the needle did not change its magnetization from day to day.

Enter Carl Friedrich Gauss (1777-1855). At Gottingen University he was involved in mapping the geomagnetic field, as had many others before. Recognizing the need for an absolute measure, in 1832 he devised a way to do just that. It was the start of real units in electromagnetism; his method was extended and grew to be the foundation for what became called the ***Electromagnetic Units*** system, the first system devised.

### GAUSS' METHOD

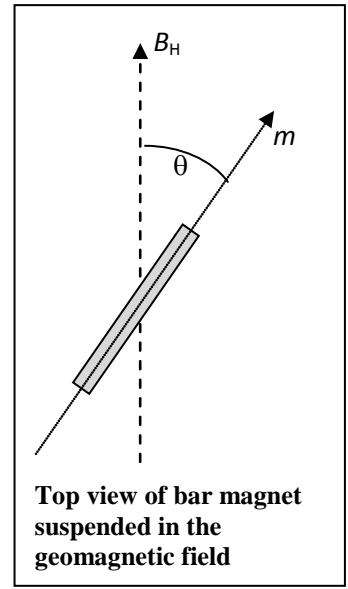

**Top view of bar magnet suspended in the geomagnetic field**

As was common, Gauss first determined the declination and inclination of the field, and then suspended a bar magnetic from a thin fiber with negligible restoring force, leaving the mass free to rotate, but constrained to rotate in the horizontal plane. Its magnetic moment **m** is unknown, and the strength $B$ of the geomagnetic field **B**, and its horizontal component $B_H$, are also unknown. A top view of the arrangement is sketched in the figure.

**m** interacts with **B** with an energy

$$U = -\mathbf{m}\cdot\mathbf{B} = -mB_H \cos\theta ,$$

assuming **m** is horizontal (in this Appendix we use SI units; Gauss of course did not).

**m** will stop oscillating when it is aligned with $B_H$. When not aligned, let θ be the angle between **m** and $B_H$ measured in the horizontal plane. The magnetic torque on **m** is

$$\tau = -\partial U/\partial\theta = -mB_H\sin\theta = -mB_H\theta \text{ for small } \theta.$$

This torque is also $\tau = J d^2\theta/dt^2$, where $J$ is the moment of inertia, so the needle oscillates as $\theta = \theta_o \sin \omega t$, where $\omega = \sqrt{(mB_H/J)}$. Thus by measuring $J$ and $\omega$ (or the oscillation period $T = 2\pi/\omega$), one has determined the product

$$mB_H = \omega^2 J$$

in terms of mechanical quantities. This determines the product of $m$ and $B_H$ in absolute terms, but not either factor separately.

The defining moment in the history of units in electromagnetism is when Gauss thought of a separate way to measure the ratio $m/B_H$. With the ratio, and the previous product, one can determine both $m$ and $B_H$ (and so $B$) in absolute terms.

## MEASURING THE RATIO $m/B_H$

Now consider the magnetic field created by the bar magnet itself. Remove the bar magnet from the fiber and mount it rigidly perpendicular to $B_H$. Let the $y$ axis be parallel to $B_H$, and $x$ parallel to $\boldsymbol{m}$. The magnetic field $B_m$ of $\boldsymbol{m}$ is the usual dipole field.

Place a second magnetic moment $m_2$, say a compass needle, on the axis of $\boldsymbol{m}$ a distance $R$ from it. Its physical size should be small compared with the distances over which $B_m$ varies. $R$ should be large compared with the dimensions of the bar magnet; however corrections can easily be made.

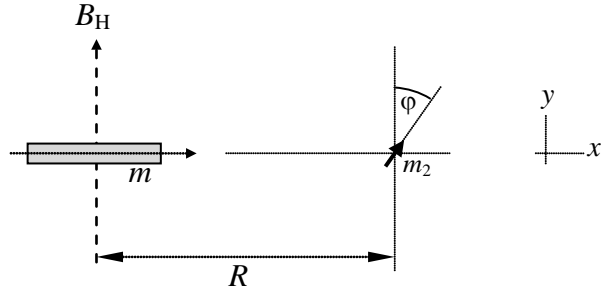

$m_2$ responds to the local total field. From $B_H$ it feels a torque

$$\tau_E = -m_2 B_H \sin\varphi$$

tending to align $m_2$ with $B_H$. And from $\boldsymbol{m}$ it feels a torque

$$\tau_m = m_2 B_m \cos\varphi$$

tending to align $m_2$ with $\boldsymbol{m}$. In equilibrium, $m_2$ will end up pointing at some angle $\varphi$ relative to $B_H$ such that

$$\tau_E = -\tau_m, \quad \text{or}$$

$$B_m = B_H \tan\varphi .$$

On the $x$ axis, the field of the bar magnet is $B_m = (\mu_o/4\pi)2m/R^3$, and so

$$m/B_H = (4\pi/\mu_o)\, \tfrac{1}{2} R^3 \tan\varphi \;.$$

Thus we have the ratio of *m* to $B_H$. Together with the previous equation for the product $mB_H$, we have

$$B_H = \sqrt{\frac{\mu_o}{4\pi}\frac{2\omega^2 J}{R^3 \tan\varphi}} \tag{A.1}$$

And so the magnetic field is determined in terms of mechanical quantities mass, length, and time.

The magnetic moment is found to be

$$m = \sqrt{\frac{4\pi}{\mu_o}\omega^2 J\, \tfrac{1}{2} R^3 \tan\varphi} \tag{A.2}$$

Gauss' method also provided the first measure of a magnetic moment in absolute terms.

Gauss' experiment has been widely discussed [e.g., Le43, Ha49, Be64, Fe68]. It has been extended and reproduced for pedagogical purposes [Va13].

# 10 APPENDIX B

# APPLYING CONVERSION RELATIONS

In mechanics, involving only mass, length, and time, the conversion of one system of units to another is trivial. A physical quantity has the same *dimensions* in all systems (English, metric, or other). For example, in converting length,

$$1 \text{ light-year} = 9.4605 \times 10^{17} \text{ cm} = 9.4605 \times 10^{15} \text{ m} = 1.0346 \times 10^{16} \text{ yards} = \ldots .$$

All sides of a conversion equation have the same dimensions, here length, and the *units* are specified. The conversion factor from one system of units to another is a dimensionless number.

But, as mentioned, in electromagnetism a physical quantity (charge, current, voltage, resistance, …) may have different dimensions in different systems. The conversion factor for a given physical quantity from one system of units (say, Gaussian) to another (say, SI) itself has dimensions. This fact causes confusion and can produce subtle errors.

### EXAMPLE 1

There is a commonly confusing point about the ratio of the emu unit of charge to the esu unit of charge. The same argument applies to the ratio of currents.

As put by Maxwell[60] if a charge is measured to contain 1 unit of charge in electromagnetic units, i.e., 1 abC, it contains $3 \times 10^{10}$ units of charge in electrostatic units, i.e., $3 \times 10^{10}$ stC. This relation is commonly written 1 abC = $3 \times 10^{10}$ stC. Since the $3 \times 10^{10}$ derives from the speed of light, it appears to have dimensions L/T. It is commonly stated: the ratio of emu unit of charge to esu unit of charge is the speed of light.

Let us check the statement that 1 abC = $3 \times 10^{10}$ stC by appealing to the definitions of charge in emu and esu, a procedure which would appear to be the safest of methods. Referring directly to the definitions of 1 abC and 1 stC, Eqs (29) and (50), find

$$\frac{1 \text{ abC}}{1 \text{ stC}} = \frac{1 \text{ dyn}^{1/2} \cdot \text{sec}}{1 \text{ dyn}^{1/2} \cdot \text{cm}} = \frac{1 \text{ sec}}{1 \text{ cm}} = \frac{1}{\text{velocity}} \tag{B.1}$$

But this has just the inverse of the expected dimensions. And its value is 1 sec/cm, not $3 \times 10^{10}$ cm/sec. The result is wildly incorrect. What is wrong with this line of reasoning?

Referring to how many esu units of charge make up 1 emu unit, one tends to frame Maxwell's observation as an equation

$$1 \text{ abCoulomb} = 3 \times 10^{10} \text{ stCoulomb} \tag{B.2}$$

However this is not a proper "equation". In a proper equation the dimensions of the left and right sides must agree. In (B.2) the left side does not have the same dimensions as the right; the left side is in emu ($\mathrm{dyn}^{1/2} \cdot \mathrm{sec}$), and the right is in esu ($\mathrm{dyn}^{1/2} \cdot \mathrm{cm}$). For example, inserting definitions in terms of basic dimensions (1 abC = 1 $\mathrm{g}^{1/2} \cdot \mathrm{cm}^{1/2}$; 1 stC = 1 $\mathrm{g}^{1/2} \cdot \mathrm{cm}^{3/2} \cdot \mathrm{s}^{-1}$), and treating (B.2) as a normal equation, one would conclude that

$$1\,\mathrm{g}^{1/2}\mathrm{cm}^{1/2} = 3\times 10^{10}\,\mathrm{g}^{1/2}\mathrm{cm}^{3/2}\mathrm{s}^{-1}, \qquad \text{or}$$
$$1 = 3\times 10^{10}\,\mathrm{cm\,s}^{-1} \tag{B.3}$$

which is patent nonsense.

Rather, the equal sign should be replaced by "corresponds to", or, better yet, the statement should read "a charge found to be 1 abC in emu will be found to be $3\times 10^{10}$ stC in esu".

To emphasize the correspondence, as opposed to the equality, between abC and stC, the IEC has used[72] the notational symbol $\triangleq$ to mean "corresponds to", so that instead of (B.2) one writes

$$1\,\mathrm{abCoulomb} \triangleq 3\times 10^{10}\,\mathrm{stCoulomb}. \tag{B.4}$$

Consider a given charge $Q$. When expressed in emu $Q$ is a certain number of emu charge units:

$$Q = x_{emu}\,\mathrm{abC}, \qquad \text{(emu)}$$

where "abC" is the emu charge unit, and $x_{emu}$ is a dimensionless number telling how many emu units of charge are in $Q$.

If done in esu, the same charge would be written

$$Q = x_{esu}\,\mathrm{stC}, \qquad \text{(esu)}$$

where "stC" is the esu charge unit and $x_{esu}$ is a dimensionless number telling how many esu units of charge are in $Q$. Although they represent the same charge, the two expressions for $Q$ are still not "equal", for one is in emu and the other in esu, and they have different dimensions.

The safe way to compare the size of a unit in one electromagnetic system with that in another system is through the expressions for a single *mechanical* quantity expressed in the two systems, because dimensions will then be the same. If two charges $Q$ are $r$ apart, Eq (30) tells us the force between them, written in emu:

$$f(\mathrm{dyn}) = c^2 \frac{Q^2}{r^2} = c^2 \frac{(x_{emu}\,\mathrm{abC})^2}{r^2} \tag{B.5}$$

while Eq (49) is the same force between them written in esu:

$$f = \frac{Q^2}{r^2} = \frac{(x_{esu}\,\mathrm{stC})^2}{r^2}. \tag{B.6}$$

---

[72] For example, IEC Report 80000-6 (2008); email from Joanna Goodwin of the IEC, April 16, 2014. The importance of "corresponds to" rather than "equals" has been emphasized by Page [Pa70a].

(B.5) and (B.6) are equal (same mechanical dimensions), obtaining

$$x_{esu} \text{ stC} = c \, x_{emu} \text{ abC} . \tag{B.7}$$

This is a valid equation in dimensions and magnitudes, since the dimensions of stC are velocity times the dimensions of abC.

So if $Q$ is 1 abC, $x_{emu} = 1$, and $x_{esu} =$ (the number of electrostatic units of charge to equal 1 electromagnetic unit of charge) $= c = 3 \times 10^{10}$ cm/sec. This demonstrates Maxwell's statement in footnote 60, the "conversion factor" from emu charge to esu charge has dimensions of velocity. If the "cm/sec" is separated from $c$ and considered to multiply 1 abC, the product is 1 stC, as follows from Eq (B.1), allowing (B.7) to be interpreted as $3 \times 10^{10}$ stC = 1 abC. Either way, the "cm/sec" must be included.

A bucket of $10^{20}$ electrons has a charge of 1.602 abCoulomb, or 16.02 Coulomb, or $4.803 \times 10^{10}$ statCoulomb.

### EXAMPLE 2

A second example of the danger of directly applying the familiar transitive law of equality (if A = B, and B = C, then A = C) to conversion factors crops up in a simple exercise.

From the definition of an abAmpere in emu we have

$$1 \text{ abA} = (1 \text{ dyn})^{1/2} \qquad \text{[correct equation; it defines the abA, Eq (28)]}$$

But we also know that relative to SI,

$$1 \text{ abA} = 10 \text{ A} \qquad \text{[a conversion factor, the = sign should be replaced by } \hat{=}, \text{ or "corresponds to"]}$$

If one were to combine these two relations as equations, one would obtain
$10 \text{ A} = 1 \text{ abA} = (1 \text{ dyn})^{1/2}$, or $1 \text{ A} = 0.1 \text{ dyn}^{1/2}$, a relation which is not at all true.

By treating conversion relations as equations, one can construct nonsensical relations.

### EXAMPLE 3

As a third example, consider the well known conversion of the unit of $B$ from SI (tesla T) to emu or Gaussian (gauss):

$$1 \text{ T} = 10^4 \text{ gauss} .$$

One can attempt to obtain the definition of 1 gauss in emu or Gaussian units by proceeding as follows from this conversion factor,

$$10^4 \text{ gauss} = 1 \text{ T} = 1 \text{ Wb/m}^2 = 1 \text{ V·s/m}^2 = 1 \text{ J·s/C·m}^2 = 1 \text{ J/A·m}^2 = 10^7 \text{ erg/A·m}^2$$
$$= 10^8 \text{ erg/abA·m}^2 = 10^4 \text{ erg/abA·cm}^2 = 10^4 \text{ dyn/abA·cm} = 10^4 \text{ abA/cm},$$

where we used 1 abA = 10 A and 1 abA = 1 dyn$^{1/2}$.

Or, dividing by $10^4$,

$$1 \text{ gauss} = 1 \text{ abA/cm}.$$

This result is correct and reproduces the definition of Section 6.1.1.

But now, rather than with the unit of **B**, try the same conversion for the unit of **H**, from the relation of an oersted relative to SI:

$$1 \text{ oersted} = 1000/4\pi \text{ A/m}.$$

Since 1 abA = 10 A, and 1 m = 100 cm, one would similarly conclude that

$$1 \text{ oersted} = 10/4\pi \text{ A/cm} = 1/4\pi \text{ abA/cm} \quad \text{(wrong)}$$

But in fact,

$$1 \text{ oersted} = 1 \text{ abA/cm} \quad \text{(correct)}$$

from Section 6.1.2. What is correct is that 1 oersted $\triangleq$ 1000/4π A/m. One cannot reliably "back convert" from the oersted in terms of SI quantities to the oersted in emu by using other conversion factors like 10 A = 1 abA, since the relation 1 Oe = 1000/4π A/m is a conversion factor, not a bonafide equality. In view of the simplicity of the procedure, this is a particularly blatant example.[73]

In the 19$^{th}$ century, the oersted was defined as the magnetic field strength 1 cm from a unit magnetic pole. Today a common definition is the field that exerts a force of $f' = 1$ dyn/cm on a wire carrying current 1 abA, $f' = IH$.

The best way (least error prone way) to compare units, and to define 1 Oe, is again via a physical setup, equating a mechanical quantity analyzed in two unit systems. Consider a wire carrying current $I$ in a field $H$. In emu, where $B = H$, the force per unit length on the wire is

$$f' = IB = IH, \quad \text{or}$$

$$f'(\text{dyn/cm}) = I(\text{abA})B(\text{gauss}) = I(\text{abA})H(\text{Oe}).$$

In SI the force is

$$f'(\text{N/m}) = IB = I\mu_o H = I(\text{A})\mu_o(\text{H/m})H(\text{A/m}).$$

These two mechanical forces may be equated, and, using $f'(\text{N/m}) = 10^{-3} f'(\text{dyn/cm})$, obtain

$$f'(\text{N/m}) = \mu_o(\text{H/m})I(\text{A})H(\text{A/m}) = 4\pi \times 10^{-7} I(\text{A})H(\text{A/m}) = 10^{-3} f'(\text{dyn/cm})$$

$$= 10^{-3} I(\text{abA})H(\text{Oe}), \quad \text{or}$$

$$4\pi \times 10^{-4} I(\text{A})H(\text{A/m}) = I(\text{abA})H(\text{Oe}).$$

And since a current measured to be $I(\text{A})$ will be measured to be $10I(\text{abA})$, we have

$$4\pi \times 10^{-3} I(\text{abA})H(\text{A/m}) = I(\text{abA})H(\text{Oe}), \quad \text{or}$$

---

[73] In the units appendix of Condon & Odishaw, eds., *Handbook of Physics* (1$^{st}$ ed. 1958 [Co58]; 2$^{nd}$ ed. 1967) it states 1 oersted = 1/4π abA/cm. This agreement with the (incorrect) back-conversion above misled the present writer for a long time. On closer examination the Handbook article appears to contain statements that are not obviously consistent, and the present writer found it difficult to follow. The writer is indebted to Prof. Kirk McDonald of Princeton University for many helpful email exchanges on the definition of 1 oersted. See his note concerning *D* and *H* [Mc14].

$$4\pi \times 10^{-3} H(\text{A/m}) = H(\text{Oe}),$$

which is the usual, correct, conversion relation 1 Oe $\triangleq 10^3/4\pi$ A/m.

The parallel argument in terms of $B$ instead of $H$, reproduced the correct 1 gauss $\triangleq 10^{-4}$ T. A factor $4\pi$ enters in the $H$ argument because $\mu_o$ enters the SI side but not the Gaussian side. But if one considers 1 Oe $= 10^3/4\pi$ A/m to be an equation, and "back converts" by replacing 1 A by 0.1 abA, the $4\pi$ remains and one ends up with the wrong definition of 1 Oe.

The same correct $H$ conversion between Gaussian and SI units can be obtained more simply as follows:

$$B(\text{T}) = \mu_o H(\text{A/m}) = 10^{-4} B(\text{gauss}) = 10^{-4} H(\text{Oe}), \text{ so that}$$

$$H(\text{Oe}) = \mu_o \times 10^4 H(\text{A/m}) = 4\pi \times 10^{-3} H(\text{A/m}).$$

But it is safer in all cases to revert to a physical setup analyzed in two systems.

Rather than the force on a wire, a similar correct "conversion" argument can be based on the field inside a long solenoid, or the field created by a wire at radius $r$. To obtain the correct conversion relation it is safest to analyze a well defined physical setup, and compare a mechanical quantity (like a force or energy) in two unit systems.

### EXAMPLE 4

A fourth example can be taken from the conversion of capacitance from Gaussian units to SI. In Gaussian units (or in esu), capacitance $C$ (charge/potential) has base dimensions of length, and unit of cm. The unit may be called 1 *StFarad* = 1 cm. In SI capacitance has dimensions [F] = Coulomb/Volt, or $A^2 \cdot \text{sec}^4/\text{kg} \cdot \text{m}^2$.

Again, the reliable procedure is to set up a physical problem and express the desired quantity in both systems. Consider a parallel plate capacitor in vacuum with plates of area $A$ separated by distance $d$. A charge $+/- Q$ resides on the inner surface of each plate.

Obtain the voltage by applying Gauss' Law to a small pillbox enclosing a patch of the $+Q$ surface to obtain the electric field, then multiply by $d$.

In Gaussian units, the law reads $\nabla \cdot E = 4\pi\rho$. Integrating over a small pillbox, the electric field is found to be $E = 4\pi Q/A$, and the voltage is $V(\text{stV}) = 4\pi Q(\text{stC}) d(\text{cm})/A(\text{cm}^2)$. The capacitance is

$$C(\text{cm}) = Q/V = A(\text{cm}^2)/4\pi d(\text{cm}).$$

In SI the same capacitance, now in Farads, appears as

$$C(\text{F}) = \epsilon_o \frac{A}{d} = \frac{1}{\mu_o c^2} \frac{A(\text{m}^2)}{d(\text{m})} = \frac{1}{4\pi \times 10^{-7} \cdot (3 \times 10^8)^2} 0.01 \frac{A(\text{cm}^2)}{d(\text{cm})}$$

$$= \left(\frac{0.01}{9 \times 10^9}\right) \frac{A(\text{cm}^2)}{4\pi d(\text{cm})} = \frac{1}{9 \times 10^{11}} C(\text{cm})$$

Thus it takes $9 \times 10^{11}$ Gaussian units of capacitance to equal 1 F. Or the *unit* 1 F and the unit 1 cm are related by

$$1 \text{ F} \triangleq 9 \times 10^{11} \text{ cm.}$$

The esu and Gaussian unit of capacitance, 1 cm, is very small, roughly 1 pF.

We might mention that in this example, simple substitution from conversion relations happens to work. The unit of capacitance in Gaussian units or in esu is

$$1 \text{ Gaussian unit of capacitance} = \frac{1 \text{ Gaussian unit of charge}}{1 \text{ Gaussian unit of voltage}} = \frac{1 \text{ stC}}{1 \text{ stV}} = 1 \text{ cm.} \tag{B.8}$$

Then, substituting 1 stC $\triangleq 1/(3 \times 10^9)$ C, and 1 stV $\triangleq$ 300 V as equations, obtain

$$1 \text{ cm} \triangleq \frac{1}{3 \times 10^9 \times 300} \frac{\text{C}}{\text{V}} = \frac{1}{9 \times 10^{11}} \text{F} \tag{B.9}$$

But, as mentioned, one must not rely on the procedure in general.

## 11  APPENDIX C
**EARLY HINT OF A CONNECTION BETWEEN ELECTRICITY (LIGHTNING) AND MAGNETISM**
P. Dod, *Phil. Trans. Roy. Soc.* **39**, p. 74 (1 January 1735).

VIII. *An Account of an extraordinary Effect of Lightning in communicating Magnetism. Communicated by* Pierce Dod, *M. D. F. R. S. from Dr.* Cookson *of* Wakefield *in* Yorkshire.

A Tradesman in this Place having put up a great Number of Knives and Forks in a large Box, some in Cases or Sheaths, and others not, of different Sizes, and of different Persons making, in order to be sent beyond Sea; and having placed the Box in the Corner of a large Room, there happen'd a sudden Storm of Thunder, Lightning, &c. by which the Corner of the Room was damaged, the Box split, and a good many Knives and Forks melted, the Sheaths being untouch'd. The Owner emptying the Box upon a Counter where some Nails lay, the Persons who took up the Knives, that lay upon the Nails, observed that the Knives took up the Nails. Upon this the whole Numbers was try'd, and found to do the same, nay, to such a degree as to take up large Nails, Packing-Needles, and other Iron Things of considerable Weight. Needles or other Things placed upon a Pewter-Dish, would follow the Knife or Fork, though held under the Dish, and would move along as the Knife or Fork was moved; with several other odd Appearances, which I won't now trouble you with, only this, that though you heat the Knives red-hot, yet their Power is still the same when cold.

You may be assur'd of the Truth of this, having myself made a good many Trials of the Knives and the Forks: How they came by this magnetick Power, or how Lightning should be capable of communicating such a Power, is the *Quære*.

Decem. 6th,
1732.